# Bayesian Heuristics for Group Decisions

M. Amin Rahimian & Ali Jadbabaie[*]

[*]Institute for Data, Systems, and Society (IDSS), Massachusetts Institute of Technology (MIT), Cambridge, MA 02139, USA.
Emails: rahimian , jadbabai @mit.edu

We propose a model of inference and heuristic decision making in groups that is rooted in the Bayes rule but avoids the complexities of rational inference in partially observed environments with incomplete information. According to our model, the group members behave rationally at the initiation of their interactions with each other; however, in the ensuing decision epochs, they rely on a heuristic that replicates their experiences from the first stage. Subsequently, the agents use their time-one Bayesian update and repeat it for all future time-steps; hence, updating their actions using a so-called *Bayesian heuristic*. This model avoids the complexities of fully rational inference and also provides a behavioral and normative foundation for non-Bayesian updating. It is also consistent with a dual-process psychological theory of decision making, where a controlled (conscious/slow) system develops the Bayesian heuristic at the beginning, and an automatic (unconscious/fast) system takes over the task of heuristic decision making in the sequel.

We specialize this model to a group decision scenario where private observations are received at the beginning, and agents aim to take the best action given the aggregate observations of all group members. We present the implications of the choices of signal structure and action space for such agents. We show that for a wide class of distributions from the exponential family the Bayesian heuristics take the form of an affine update in the self and neighboring actions. Furthermore, if the priors are non-informative (and possibly improper), then these action updates become a linear combination. We investigate the requirements on the modeling parameters for the action updates to constitute a convex combination as in the DeGroot model. The results reveal the nature of assumptions that are implicit in the DeGroot updating and highlights the fragility and restrictions of such assumptions; in particular, we show that for a linear action update to constitute a convex combination the precision or accuracy of private observations should be balanced among all neighboring agents, requiring a notion of social harmony or homogeneity in their observational abilities. Following the DeGroot model, agents reach a consensus asymptotically. We derive the requirements on the signal structure and network topology such that the consensus action aggregates information efficiently. This involves additional restrictions on the signal likelihoods and network structure. In the particular case that all agents observe the same number of i.i.d. samples from the same distribution, then efficiency arise in degree-regular balanced structures, where all nodes listen to and hear from the same number of neighbors.

We next shift attention to a finite state model, in which agents take actions over the probability simplex; thus revealing their beliefs to each other. We show that the Bayesian heuristics, in this case, prescribe a log-linear update rule, where each agent's belief is set proportionally to the product of her own and neighboring beliefs. We analyze the evolution of beliefs under this rule and show that agents reach a consensus. The consensus belief is supported over the maximizers of a weighted sum of the log-likelihoods of the initial observations. Since the weights of the signal likelihoods coincide with the network centralities of their respective agents, these weights can be equalized in degree-regular and balanced topologies, where all nodes have the same in and out degrees. Therefore, in such highly symmetric structures the support of the consensus belief coincides with the maximum likelihood estimators (MLE) of the truth state; and here again, balanced regular structures demonstrate a measure of efficiency. Nevertheless, the asymptotic beliefs systematically reject the less probable alternatives in spite of the limited initial data, and in contrast with the optimal (Bayesian) belief of an observer with complete information of the environment and private signals. The latter would assign probabilities proportionally to the likelihood of every state, without rejecting any of the possible alternatives. The asymptotic rejection of less probable alternatives indicates a case of group polarization, i.e. overconfidence in the group aggregate that emerges as a result of the group interactions. Unlike the linear action updates and the DeGroot model which entail a host of knife-edge conditions on the signal structure and model parameters, we observe that the belief updates are unweighted; not only they effectively internalize the heterogeneity of the private observations, but also they compensate for the individual priors. Thence, we are led to the conclusion that multiplicative belief updates, when applicable provide a relatively robust description of the decision making behavior.

*Key words*: opinion dynamics; social learning; Bayesian learning; non-Bayesian learning; rational learning; observational learning; statistical learning; distributed learning; distributed hypothesis testing; distributed detection; DeGroot model; linear regression; conjugate priors; exponential families

**1. Introduction.** Daniel Kahneman in his highly acclaimed work, "*Thinking, Fast and Slow*", points out that the proper way to elicit information from a group is not through a public discussion but rather confidentially collecting each person's judgment [52, Chapter 23]. Indeed, decision making among groups of individuals exhibit many singularities and important inefficiencies that lead to Kahneman's noted advice. As a team converges on a decision expressing doubts about the wisdom of the consensus choice is suppressed; subsequently teams of decision makers are afflicted with *groupthink* as they appear to reach a consensus.[1] The mechanisms of uncritical optimism, overconfidence, and the illusions of validity in group interactions also lead to *group polarization*, making the individuals more amenable toward extreme opinions [92].

In more abstract terms, agents in a social network exchange opinions to benefit from each other's experience and information when making decisions about adoption of technologies, purchasing products, voting in elections and so on. The problem of social learning is to characterize and understand such interactions and it is a classical focus of research in microeconomics [45, Chapter 8], [39, Chapter 5], [18]. Research on formation and evolution of beliefs in social networks and subsequent shaping of the individual and group behaviors have attracted much attention amongst diverse communities in engineering [58, 97], statistics [66], economics [47, 65], and sociology [82]. An enhanced understanding of decision making and learning in social networks sheds light on the role of individuals in shaping public opinion and how they influence efficiency of information transmissions. These in turn help us improve our predictions about group behavior and provide guidelines for designing effective social and organizational policies.

**1.1. Rational social learning.** The rational approach advocates application of Bayes rule to the entire sequence of observations successively at every step. However, such repeated applications of Bayes rule in networks become very complex, especially if the agents are unaware of the global network structure; and as they use their local data to make inferences about all possible contingencies that can lead to their observations. While some analytical properties of rational learning is deduced and studies in the literature [32, 1, 68, 66, 42], their tractable modeling and analysis remains an important problem in network economics and continues to attract attention.

Some of the earliest results addressing the problem of social learning are due to Banerjee (1992) [9], and Bikhchandani et al. (1998) [13] who consider a complete graph structure where the agent's observations are public information and also ordered in time, such that each agent has access to the observations of all the past agents. These assumptions help analyze and explain the interplay between public and private information leading to fashion, fads, herds etc. Later results by Gale and Kariv [32] relax some of these assumptions by considering the agents that make simultaneous observations of only their neighbors rather than the whole network, but the computational complexities limit the analysis to networks with only two or three agents. In more recent results, Mueller-Frank [68] provides a framework of rational learning that is analytically amenable. Mossel, Sly and Tamuz [66] analyze the problem of estimating a binary state of the world from a single initial private signal that is independent and identically distributed amongst the agents conditioned on the true state; and they show that by repeatedly observing each other's best estimates of the unknown as the size of the network increases with high probability Bayesian agents asymptotically learn the true state. Hence, the agents are able to combine their initial private observations and learn the truth. Further results by Kanoria and Tamuz [53] provide efficient algorithms to calculate each agent's estimate in the case of tree networks.

On the one hand, the properties of rational learning models are difficult to analyze beyond some simple asymptotic facts such as convergence. On the other hand, these models make unrealistic assumptions about the complexity and amount of computations that agents perform before committing to a decision. To avoid these shortcomings, an alternative "*non-Bayesian*" approach relies on simple and intuitive heuristics that are descriptive of how agents aggregate the reports of their neighbors before coming up with a decision.

---

[1]Gar Klein proposes a famous method of *project premortem* to overcome the groupthink through an exercise: imagining that the the planned decision was failed in implementation and writing a brief report of the failure [57].

**1.2. Heuristic decision making.** Heuristics are used widely in the literature to model social interactions and decision making [34, 35, 36]. They provide tractable tools to analyze boundedly rational behavior and offer insights about decision making under uncertainty. Hegselmann and Krause [43] investigate various ways of averaging to model opinion dynamics and compare their performance for computations and analysis. Using such heuristics one can avoid the complexities of fully rational inference, and their suitability are also verified in experimental studies by Grimm and Mengel [40] and Chandrasekhar, Larreguy and Xandri [19]. The study of such heuristics started in 1974 with the seminal work of DeGroot [20] in linear opinion pooling, where agents update their opinions to a convex combination of their neighbors' beliefs and the coefficients correspond to the level of confidence that each agent puts in each of her neighbors. More recently, Jadbabaie, Molavi and Tahbaz-Salehi [47, 48, 65] consider a variation of this model for streaming observations, where in addition to the neighboring beliefs the agents also receive private signals. Despite their widespread applications, theoretical and axiomatic foundations of social inferences using heuristics and non-Bayesian updates have received limited attention and only recently [65, 69], and a comprehensive theory of non-Bayesian learning that reconciles the rational and boundedly rational approaches with the widely used heuristics remains in demand. The main goal of our paper is to address this gap and to do so from a behavioral perspective.

A dual process theory for the psychology of mind and its operation identifies two systems of thinking [24]: one that is fast, intuitive, non-deliberative, habitual and automatic (system one); and a second one that is slow, attentive, effortful, deliberative, and conscious (system two).[1] Major advances in behavioral economics are due to incorporation of this dual process theory and the subsequent models of bounded rationality [51]. Reliance on heuristics for decision making is a distinctive feature of system one that avoids the computational burdens of a rational evaluation; system two on the other hand, is bound to deliberate on the options based on the available information before making recommendations. The interplay between these two systems and how they shape the individual decisions is of paramount importance [17].

Tversky and Kahneman argue that humans have limited time and brainpower, therefore they rely on simple rules of thumb, i.e. heuristics, to help them make judgments. However, the use of these heuristics causes people to make predictable errors and subjects them to various biases [96]. Hence, it is important to understand the nature and properties of heuristic decision making and its consequences to individual and organizational choice behavior. This premise underlies many of the recent advances in behavioral economics [93], and it motivates our work as well.

**1.3. Our contribution.** In this work we are concerned with the operations of system one: we aim to study heuristics for information aggregation in group decision scenarios when the relevant information is dispersed among many individuals. In such situations, individuals in the group are subjected to informational (but not strategic) externalities. By the same token, the heuristics that are developed for decision making in such situations are also aimed at information aggregation. In our model as the agent experiences with her environment her initial response would engage her system two: she rationally evaluates the reports of her neighbors and use them to make a decision. However, after her initial experience and by engaging in repeated interactions with other group members her system one takes over her decision processes, implementing a heuristic that imitates her (rational/Bayesian) inferences from her initial experience; hence avoiding the burden of additional cognitive processing in the ensuing interactions with her neighbors.

---

[1] While many decision science applications focus on developing dual process theories of cognition and decision making (cf. [25, 63] and the references therein); other researchers identify multiple neural systems that derive decision making and action selection: ranging from reflexive and fast (Pavlovian) responses to deliberative and procedural (learned) ones; and these systems are in turn supported by several motoric, perceptual, situation-categorization and motivational routines which together comprise the decision making systems [80, Chapyter 6].

On the one hand, our model of inference based on Bayesian (initial) heuristics is motivated by the real-world behavior of people induced by their system one and reflected in their spur-of-the-moment decisions and impromptu behavior: basing decisions only on the immediately observed actions and disregarding the history of the observed actions or the possibility of correlations among different observations; i.e. "*what you see is all there is*" [52]. On the other hand, Bayesian heuristics offer a boundedly rational approach to model decision making over social networks. The latter is in the sense of the word as coined by Herbert A. Simon: to incorporate modifications that lead to substantial simplifications in the original choice problem [87].[1] This in contrast with the Bayesian approach which is not only unrealistic in the amount of cognitive burden that it imposes on the agents, but also is often computationally intractable and complex to analyze.

Our main contribution is to offer a normative framework for heuristic decision making, by relying on the time-one Bayesian update and using it for all future decision epochs. This model offers a behavioral foundation for non-Bayesian updating that is compatible with the dual-process psychological theory of decision making. In Section 2, we describe the mathematical details of our model; in particular, we explain the mathematical steps for deriving the so-called Bayesian heuristics in a given decision scenario. Specific cases of Bayesian heuristics that we explore in the following sections are the log-linear (multiplicative) updating of beliefs over the probability simplex, and the linear (weighted arithmetic average) updating of actions over the Euclidean space. In Section 3, we specialize our group decision model to a setting involving exponential family of distributions for both signal likelihoods and agents' beliefs. The agents aim to estimate the expected values of the sufficient statistics for their signal structures. We show that the Bayesian heuristics in this case are affine rules in the self and neighboring actions, and we give explicit expressions for their coefficients. Subsequently, we provide conditions under which these action updates constitute a convex combination as in the DeGroot model, with actions converging to a consensus in the latter case. We also investigate the efficiency of the consensus action in aggregating the initial observations of all agents across the network. Next in Section 4, we discuss a situation where agents exchange beliefs about a truth state that can takes one of the finitely many possibilities. The Bayesian heuristics in this case set the updated beliefs proportional to the product of self and neighboring beliefs, as reported in every decision epoch. We investigate the evolution of beliefs under the prescribed update rules and compare the asymptotic beliefs with that of a Bayesian agent with direct access to all the initial observations; thus characterizing the inefficiencies of the asymptotic beliefs. We summarize our findings from the analysis of linear action and log-linear belief updates in Section 5, where we reiterate the assumptions that are implicit in the adoption of popular aggregation heuristics such as the DeGroot model; moreover, we discuss the inefficiencies that arise as a result of their application. Such heuristics allow us to aggregate the information in our environment, and provide for desirable asymptotic properties such as consensus; however, this consensus often fails as an efficient group aggregate for the individuals' private data. We provide the mathematical proofs and the relevant details for many of the results in the appendices at end of the paper.

---

[1]Simon advocates "bounded rationality" as compatible with the information access and the computational capacities that are actually possessed by the agents in their environments. Most importantly he proposes the use of so-called "*satisficing*" heuristics; i.e. to search for alternatives that exceed some "*aspiration levels*" by satisfying a set of minimal acceptability criteria [88, 89].

**2. The model.** Consider a group of $n$ agents that are labeled by $[n]$ and interact according to a digraph $\mathcal{G} = ([n], \mathcal{E})$.[1] The neighborhood of agent $i$ is the set of all agents whom she observes including herself, and it is denoted by $\mathcal{N}_i = \{j \in [n]; (j, i) \in \mathcal{E}\} \cup \{i\}$; every node having a self-loop: $(i, i) \in \mathcal{E}$ for all $i$. We refer to the cardinality of $\mathcal{N}_i$ as the degree of node $i$ and denote it by $deg(i)$. There is a state $\theta \in \Theta$ that is unknown to the agents and it is chosen arbitrarily by nature from an underlying state space $\Theta$, which is measurable by a $\sigma$-finite measure $\mathcal{G}_\theta(\cdot)$. For example if a space ($\Theta$ or $\mathcal{S}$) is a countable set, then we can take its $\sigma$-finite measure ($\mathcal{G}_\theta$ or $\mathcal{G}_s$) to be the counting measure, denoted by $\mathcal{K}(\cdot)$; and if the space is a subset of $\mathbb{R}^k$ with positive Lebesgue measure, then we can take its $\sigma$-finite measure to be the Lebesgue measure on $\mathbb{R}^k$, denoted by $\Lambda_k(\cdot)$. Associated with each agent $i$, $\mathcal{S}_i$ is a measurable space called the signal space of $i$, and given $\theta$, $\mathcal{L}_i(\cdot \mid \theta)$ is a probability measure on $\mathcal{S}_i$, which is referred to as the *signal structure* of agent $i$. Furthermore, $(\Omega, \mathscr{F}, \mathcal{P}_\theta)$ is a probability triplet, where $\Omega = \mathcal{S}_1 \times \ldots \times \mathcal{S}_n$ is a product space, and $\mathscr{F}$ is a properly defined sigma field over $\Omega$. The probability measure on $\Omega$ is $\mathcal{P}_\theta(\cdot)$ which assigns probabilities consistently with the signal structures $\mathcal{L}_i(\cdot \mid \theta), i \in [n]$; and in such a way that with $\theta$ fixed, the random variables $\mathbf{s}_i$, $i \in [n]$ taking values in $\mathcal{S}_i$, are independent. These random variables represent the private signals that agents $i \in [n]$ observe at time 0. Note that the private signals are independent across the agents. The expectation operator $\mathbb{E}_\theta\{\cdot\}$ represents integration with respect to $\mathcal{P}_\theta(d\omega)$, $\omega \in \Omega$.

**2.1. Beliefs, actions and rewards.** An agents' belief about the unknown allows her to make decisions even as the outcome is dependent on the unknown value $\theta$. These beliefs about the unknown state are probability distributions over $\Theta$. Even before any observations are made, every agent $i \in [n]$ holds a prior belief $\mathcal{V}_i(\cdot) \in \Delta\Theta$; this represents her subjective biases about the possible values of $\theta$. For each time instant $t$, let $\boldsymbol{\mathcal{M}}_{i,t}(\cdot)$ be the (random) probability distribution over $\Theta$, representing the *opinion* or *belief* at time $t$ of agent $i$ about the realized value of $\theta$. Moreover, let the associated expectation operator be $\mathbb{E}_{i,t}\{\cdot\}$, representing integration with respect to $\boldsymbol{\mathcal{M}}_{i,t}(d\theta)$. We assume that all agents share the common knowledge of signal structures $\mathcal{L}_i(\cdot|\hat{\theta}), \forall \hat{\theta} \in \Theta$, their priors $\mathcal{V}_i(\cdot)$, and their corresponding sample spaces $\mathcal{S}_i$ and $\Theta$ for all $i \in [n]$. [2]

Let $t \in \mathbb{N}_0$ denote the time index; at $t = 0$ the values $\theta \in \Theta$ followed by $s_i \in \mathcal{S}_i$ of $\mathbf{s}_i$ are realized and the latter is observed privately by each agent $i$ for all $i \in [n]$. Associated with every agent $i$ is an action space $\mathcal{A}_i$ that represents all the choices available to her at every point of time $t \in \mathbb{N}_0$, and a utility $u_i(\cdot, \cdot) : \mathcal{A}_i \times \Theta \rightarrow \mathbb{R}$ which in expectation represents her von Neumann-Morgenstern preferences regarding lotteries with independent draws from $\mathcal{A}_i$ and/or $\Theta$. These utilities are additive over time corresponding to successive independent draws. The utility functions and action spaces are common knowledge amongst the agents. Subsequently, at every time $t \in \mathbb{N}_0$ each agent $i \in [n]$ chooses an action $\mathbf{a}_{i,t} \in \mathcal{A}_i$ and is rewarded $u_i(\mathbf{a}_{i,t}, \theta)$.

---

[1] Some notations: Throughout the paper, $\mathbb{R}$ is the set of real numbers, $\mathbb{N}$ denotes the set of all natural numbers, and $\mathbb{N}_0 := \mathbb{N} \cup \{0\}$. For $n \in \mathbb{N}$ a fixed integer the set of integers $\{1, 2, \ldots, n\}$ is denoted by $[n]$, while any other set is represented by a capital Greek or calligraphic letter. For a measurable set $\mathcal{X}$ we use $\Delta\mathcal{X}$ to denote the set of all probability distributions over the set $\mathcal{X}$. Furthermore, any random variable is denoted in boldface letter, vectors are represented in lowercase letters and with a bar over them, measures are denoted by upper case Greek or calligraphic Latin letters, and matrices are denoted in upper case Latin letters. For a matrix $A$, its spectral radius $\rho(A)$ is the largest magnitude of all its eigenvalues.

[2] The signal structures $\mathcal{L}_i(\cdot|\hat{\theta}), \forall \hat{\theta} \in \Theta$ and the priors $\mathcal{V}_i(\cdot)$, as well as the corresponding sample spaces $\mathcal{S}_i$ and $\Theta$ are common knowledge amongst the agents for all $i \in [n]$. The assumption of common knowledge in the case of fully rational (Bayesian) agents implies that given the same observations of one another's beliefs or private signals distinct agents would make identical inferences; in the sense that starting form the same belief about the unknown $\theta$, their updated beliefs given the same observations would be the same; in Aumann's words, rational agents cannot agree to disagree [5].

**2.2. Aggregation heuristics.** Given $\mathbf{s}_i$, agent $i$ forms an initial Bayesian opinion $\boldsymbol{\mathcal{M}}_{i,0}(\cdot)$ about the value of $\theta$ and chooses her action $\mathbf{a}_{i,0} \hookleftarrow \arg\max_{a_i \in \mathcal{A}_i} \int_{\Theta} u_i(a_i, \hat{\theta}) \boldsymbol{\mathcal{M}}_{i,0}(d\theta)$, maximizing her expected reward. Here for a set $\mathcal{A}$, we use the notation $\mathbf{a} \hookleftarrow \mathcal{A}$ to denote an arbitrary choice from the elements of $\mathcal{A}$ that is assigned to $\mathbf{a}$. Not being notified of the actual realized value for $u_i(\mathbf{a}_{i,0}, \theta)$, she then observes the actions that her neighbors have taken. Given her extended set of observations $\{\mathbf{a}_{j,0}, j \in \mathcal{N}_i\}$ at time $t = 1$, she refines her opinion into $\boldsymbol{\mathcal{M}}_{i,1}(\cdot)$ and makes a second, and possibly different, move $\mathbf{a}_{i,1}$ according to:

$$\mathbf{a}_{i,1} \hookleftarrow \arg\max_{a_i \in \mathcal{A}_i} \int_{\Theta} u_i(a_i, \theta) \boldsymbol{\mathcal{M}}_{i,1}(d\theta), \tag{1}$$

maximizing her expected pay off conditional on everything that she has observed thus far; i.e. maximizing $\mathbb{E}_{i,1}\{u_i(a_i, \theta)\} = \mathbb{E}_{\theta}\{u_i(\mathbf{a}_{i,1}, \theta)|\mathbf{s}_i, \mathbf{a}_{j,0} : j \in \mathcal{N}_i\} = \int_{\Theta} u_i(a_i, \theta) \boldsymbol{\mathcal{M}}_{i,1}(d\theta)$. Subsequently, she is granted her net reward of $u_i(\mathbf{a}_{i,0}, \theta) + u_i(\mathbf{a}_{i,1}, \theta)$ from her past two plays. Following realization of rewards for their first two plays, in any subsequent time instance $t > 1$ each agent $i \in [n]$ observes the preceding actions of her neighbors $\mathbf{a}_{j,t-1} : j \in \mathcal{N}_i$ and takes an option $\mathbf{a}_{i,t}$ out of the set $\mathcal{A}_i$. Of particular significance in our description of the behavior of agents in the succeeding time periods $t > 1$, is the relation:

$$f_i(\mathbf{a}_{j,0} : j \in \mathcal{N}_i) := \mathbf{a}_{i,1} \hookleftarrow \arg\max_{a_i \in \mathcal{A}_i} \mathbb{E}_{i,1}\{u_i(a_i, \theta)\} \tag{2}$$

derived in (1), which given the observations of agent $i$ at time $t = 0$, specifies her (Bayesian) pay-off maximizing action for time $t = 1$. Once the format of the mapping $f_i(\cdot)$ is obtained, it is then used as a heuristic for decision making in every future epoch. The agents update their action by choosing: $\mathbf{a}_{i,t} = f_i(\mathbf{a}_{j,t-1} : j \in \mathcal{N}_i), \forall t > 1$. We refer to the mappings $f_i : \prod_{j \in \mathcal{N}_i} \mathcal{A}_j \to \mathcal{A}_i$ thus obtained, as *Bayesian heuristics.*[1,2,3]

---

[1]The heuristics thus obtained suffer from same fallacies of snap judgments that are associated with the recommendations of system one in "Thinking, Fast and Slow"; flawed judgments that rely on simplistic interpretations: "*what you see is all there is*", in Kahneman's elegant words [52]. Indeed, the use of the initial Bayesian update for future decision epochs entails a certain level of naivety on the part of the decision maker: she has to either assume that the structure of her neighbors' reports have not departed from their initial format, or that they are not being influenced back by her own or other group members and can thus be regarded as independent sources of information. Such naivety in disregarding the history of interactions has been highlighted in our earlier works on Bayesian learning without recall [75], where we interpret the use of time-one Bayesian update for future decision epochs, as a rational but memoryless behavior: by regarding their observations as being direct consequences of inferences that are made based on the initial priors, the agents reject any possibility of a past history beyond their immediate observations.

[2]Similar and related forms of naivety have been suggested in the literature. Eyster and Rabin [26, 27] propose the autarkic model of naive inference, where players at each generation observe their predecessors but naively think that any predecessor's action relies solely on that player's private information, thus ignoring the possibility that successive generations are learning from each other. Bala and Goyal [7]study another form of naivety and bounded-rational behavior by considering a variation of observational learning in which agents observe the action and pay-offs of their neighbors and make rational inferences about the action/pay-off correspondences, based on their observations of the neighboring actions; however, they ignore the fact that their neighbors are themselves learning and trying to maximize their own pay-offs. Levy and Razin look at a particularly relevant cognitive bias called correlation neglect, which makes individuals regard the sources of their information as independent [60]; they analyze its implications to diffusion of information, and focus in particular, on the voting behavior.

[3]Cognitive and psychological roots of the Bayesian heuristics as aggregation rules can be traced to Anderson's seminal theory of information integration, developed throughout 1970s and 1980s [3]. Accordingly, a so-called "*value function*" assigns psychological values to each of the stimuli and these psychological values are then combined into a single psychological (and later an observable) response through what is called the "*integration function*". A fundamental assumption is that valuation can be represented at a higher (molar) level as a value on the response dimension for each stimulus, as well as a weight representing the salience of this stimulus in the overall response. These valuations and weights are themselves the result of integration processes in the lower (molecular) level. At the heart of information integration theory is the "*cognitive algebra*" which describes the rules by which the values and weights of stimuli are integrated into an overall response [4].

**3. Affine action updates, linear updating and DeGroot learning.** In this section we explore the essential modeling features that lead to a linear structure in the Bayesian Heuristics. We present a general scenario that involves the exponential family of distributions and leads to linear action updates.

To describe the signal structures, we consider a measurable sample space $\mathcal{S}$ with a $\sigma$-finite measure $\mathcal{G}_s(\cdot)$, and a parametrized class of sampling functions $\{\mathcal{L}(\cdot|\theta; \sigma_i) \in \Delta \mathcal{S} : \sigma_i > 0 \text{ and } \delta_i > 0\}$ belonging to the $k$-dimensional exponential family as follows:

$$\ell(s|\theta; \sigma_i, \delta_i) := \frac{d\mathcal{L}(\cdot|\theta; \sigma_i, \delta_i)}{d\mathcal{G}_s} = \sigma_i \left| \frac{\Lambda_k(\xi(ds))}{\mathcal{G}_s(ds)} \right| \tau\left(\sigma_i \xi(s), \delta_i\right) e^{\sigma_i \eta(\theta)^T \xi(s) - \delta_i \gamma(\eta(\theta))}, \tag{3}$$

where $\xi(s) : \mathcal{S} \to \mathbb{R}^k$ is a measurable function acting as a sufficient statistic for the random samples, $\eta : \Theta \to \mathbb{R}^k$ is a mapping from the parameter space $\Theta$ to $\mathbb{R}^k$, $\tau : \mathbb{R}^k \times (0, +\infty) \to (0, +\infty)$ is a positive weighting function, and

$$\gamma(\eta(\theta)) := \frac{1}{\delta_i} \ln \int_{s \in \mathcal{S}} \sigma_i \left| \frac{\Lambda_k(\xi(ds))}{\mathcal{G}_s(ds)} \right| \tau(\sigma_i \xi(s), \delta_i) e^{\sigma_i \eta(\theta)^T \xi(s)} \mathcal{G}_s(ds),$$

is a normalization factor that is constant when $\theta$ is fixed, even though $\delta_i > 0$ and $\sigma_i > 0$ vary. This normalization constant for each $\theta$ is uniquely determined by the functions $\eta(\cdot)$, $\xi(\cdot)$ and $\tau(\cdot)$. The parameter space $\Theta$ and the mapping $\eta(\cdot)$ are such that the range space $\Omega_\theta := \{\eta(\theta) : \theta \in \Theta\}$ is an open subset of the natural parameter space $\Omega_\eta := \{\eta \in \mathbb{R}^k : \int_{s \in \mathcal{S}} |\Lambda_k(\xi(ds))/\mathcal{G}_s(ds)| \tau(\xi(s), 1) e^{\eta^T \xi(s)} \mathcal{G}_s(ds) < \infty\}$. In (3), $\sigma_i > 0$ and $\delta_i > 0$ for each $i$ are scaling factors that determine the quality or informativeness of the random sample $\mathbf{s}_i$ with regard to the unknown $\theta$: fixing either one of the two factors $\sigma_i$ or $\delta_i$, the value of the other one increases with the increasing informativeness of the observed value $\xi(\mathbf{s}_i)$. The following conjugate family of priors[1] are associated with the likelihood structure (3). This family is determined uniquely by the transformation and normalization functions: $\eta(\cdot)$ and $\gamma(\cdot)$, and it is parametrized through a pair of parameters $(\alpha, \beta)$, $\alpha \in \mathbb{R}^k$ and $\beta > 0$:

$$\mathcal{F}_{\gamma, \eta} := \Bigg\{ \mathcal{V}(\theta; \alpha, \beta) \in \Delta\Theta, \alpha \in \mathbb{R}^k, \beta_i > 0 :$$
$$\nu(\theta; \alpha, \beta) := \frac{d\mathcal{V}(\cdot; \alpha, \beta)}{d\mathcal{G}_\theta} = \left| \frac{\Lambda_k(\eta(d\theta))}{\mathcal{G}_\theta(d\theta)} \right| \frac{e^{\eta(\theta)^T \alpha - \beta\gamma(\eta(\theta))}}{\kappa(\alpha, \beta)},$$
$$\kappa(\alpha, \beta) := \int_{\theta \in \Theta} \left| \frac{\Lambda_k(\eta(d\theta))}{\mathcal{G}_\theta(d\theta)} \right| e^{\eta(\theta)^T \alpha - \beta\gamma(\eta(\theta))} \mathcal{G}_\theta(d\theta) < \infty \Bigg\}.$$

Furthermore, we assume that agents take actions in $\mathbb{R}^k$, and that they aim for a minimum variance estimation of the regression function or conditional expectation (given $\theta$) of the sufficient statistic $\xi(\mathbf{s}_i)$. Hence, we endow every agent $i \in [n]$ with the quadratic utility $u_i(a, \theta) = -(a - m_{i,\theta})^T(a - m_{i,\theta})$, $\forall a \in \mathcal{A}_i = \mathbb{R}^k$, where $m_{i,\theta} := \mathbb{E}_{i,\theta}\{\xi(\mathbf{s}_i)\} := \int_{s \in \mathcal{S}} \xi(s) \mathcal{L}(ds|\theta; \sigma_i, \delta_i) \in \mathbb{R}^k$.

Our main result in this section prescribes a scenario in which each agent starts from a prior belief $\mathcal{V}(\cdot; \alpha_i, \beta_i)$ belonging to $\mathcal{F}_{\gamma, \eta}$ and she observes a fixed number $n_i$ of i.i.d. samples from the distribution $\mathcal{L}(\cdot | \theta; \sigma_i, \delta_i)$. The agents then repeatedly communicate their actions aimed at minimum variance estimation of $m_{i,\theta}$. These settings are formalized under the following assumption that we term the *Exponential Family Signal-Utility Structure*.

---

[1] Consider a parameter space $\Theta$, a sample space $\mathcal{S}$, and a sampling distribution $\mathcal{L}(\cdot|\theta) \in \Delta\mathcal{S}$, $\theta \in \Theta$. Suppose that $\mathbf{s}$ is a random variable which is distributed according to $\mathcal{L}(\cdot|\theta)$ for any $\theta$. A family $\mathcal{F} \subset \Delta\Theta$ is a conjugate family for $\mathcal{L}(\cdot|\theta)$, if starting from any prior distribution $\mathcal{V}(\cdot) \in \mathcal{F}$ and for any signal $s \in \mathcal{S}$, the posterior distribution given the observation $\mathbf{s} = s$ belongs to $\mathcal{F}$.

ASSUMPTION 1 (**Exponential family signal-utility structure**).

(i) *Every agent $i \in [n]$ observes $n_i$ i.i.d. private samples $\mathbf{s}_{i,p}, p \in [n_i]$ from the common sample space $\mathcal{S}$ and that the random samples are distributed according to the law $\mathcal{L}(\cdot | \theta; \sigma_i, \delta_i)$ given by (3) as a member of the $k$-dimensional exponential family.*

(ii) *Every agent starts from a conjugate prior $\mathcal{V}_i(\cdot) = \mathcal{V}(\cdot; \alpha_i, \beta_i) \in \mathcal{F}_{\gamma,\eta}$, for all $i \in [n]$.*

(iii) *Every agent chooses actions $a \in \mathcal{A}_i = \mathbb{R}^k$ and bears the quadratic utility $u_i(a, \theta) = -(a - m_{i,\theta})^T (a - m_{i,\theta})$, where $m_{i,\theta} := \mathbb{E}_{i,\theta}\{\xi(\mathbf{s}_i)\} := \int_{s \in \mathcal{S}} \xi(s) \mathcal{L}(ds | \theta; \sigma_i, \delta_i) \in \mathbb{R}^k$.*

The Bayesian heuristics $f_i(\cdot), i \in [n]$ under the settings prescribed by the exponential family signal-utility structure (Assumption 1) are linear functions of the neighboring actions with specified coefficients that depend only on the likelihood structure parameters: $n_i$, $\sigma_i$ and $\delta_i$ as well as the prior parameters: $\alpha_i$ and $\beta_i$, for all $i \in [n]$.[1]

THEOREM 1 (**Affine action updates**). *Under the exponential family signal-utility structure specified in Assumption 1, the Bayesian heuristics describing the action update of every agent $i \in [n]$ are given by: $\mathbf{a}_{i,t} = f_i(\mathbf{a}_{j,t-1} : j \in \mathcal{N}_i) = \sum_{j \in \mathcal{N}_i} T_{ij} \mathbf{a}_{j,t-1} + \epsilon_i$, where for all $i, j \in [n]$ the constants $T_{ij}$ and $\delta_i$ are as follows:*

$$T_{ij} = \frac{\delta_i \sigma_j (n_j + \delta_j^{-1} \beta_j)}{\sigma_i (\beta_i + \sum_{p \in \mathcal{N}_i} n_p \delta_p)}, \; \epsilon_i = -\frac{\delta_i}{\sigma_i (\beta_i + \sum_{p \in \mathcal{N}_i} n_p \delta_p)} \sum_{j \in \mathcal{N}_i \setminus \{i\}} \alpha_j.$$

The action profile at time $t$ is the concatenation of all actions in a column vector: $\bar{\mathbf{a}}_t = (\mathbf{a}_{1,t}^T, \ldots, \mathbf{a}_{n,t}^T)^T$. The matrix $T$ with entries $T_{ij}$, $i, j \in [n]$ given in Theorem 1 is called the social influence matrix. The constant terms $\epsilon_i$ in this theorem appear as the rational agents attempt to compensate for the prior biases of their neighbors when making inferences about the observations in their neighborhood; we denote $\bar{\epsilon} = (\epsilon_1^T, \ldots, \epsilon_n^T)^T$ and refer to it as the vector of neighborhood biases. The evolution of action profiles under conditions of Theorem 1 can be specified as follows: $\bar{\mathbf{a}}_{t+1} = (T \otimes I_k) \mathbf{a}_t + \bar{\epsilon}$, where $I_k$ is the $k \times k$ identity matrix and $(T \otimes I_k)$ is a Kronecker product. Subsequently, the evolution of action profiles over time follows a non-homogeneous positive linear discrete-time dynamics, cf. [28]. If the spectral radius of $T$ is strictly less than unity: $\rho(T) < 1$, then $I - T$ is non-singular; there is a unique equilibrium action profile given by $\bar{a}_e = ((I - T)^{-1} \otimes I_k) \bar{\epsilon}$ and $\lim_{t \to \infty} \bar{\mathbf{a}}_t = \bar{a}_e$. If unity is an eigenvalue of $T$, then there may be no equilibrium action profiles or an infinity of them. If $\rho(T) > 1$, then the linear discrete-time dynamics is unstable and the action profiles may grow unbounded in their magnitude, cf. [49].

EXAMPLE 1 (**Gaussian Signals with Gaussian Beliefs**). Mossel and Tamuz [67] consider the case where the initial private signals as well as the unknown states are normally distributed and the agents all have full knowledge of the network structure. They show that by iteratively observing their neighbors' mean estimates and updating their beliefs using Bayes rule all agents converge to the same belief. The limiting belief is the same as what a Bayesian agent with direct access to everybody's private signals would have hold; and furthermore, the belief updates at each step can be computed efficiently and convergence occurs in a number of steps that is bounded in the network size and its diameter. These results however assume complete knowledge of the network structure by all the agents.

Here, we consider the linear action updates in the Gaussian setting. Let $\Theta = \mathbb{R}$ be the parameter space associated with the unknown parameter $\theta \in \Theta$. Suppose that each agent $i \in [n]$ holds onto a Gaussian prior belief with mean $\alpha_i \beta_i^{-1}$ and variance $\beta_i^{-1}$; here, $\gamma(\theta) = \theta^2/2$ and $\eta(\theta) = \theta$. Further suppose that each agent observes an independent private Gaussian signal $\mathbf{s}_i$ with mean $\theta$ and variance $\sigma_i^{-1} = \delta_i^{-1}$, for all $i \in [n]$; hence, $\xi(\mathbf{s}_i) = \mathbf{s}_i$ and $\tau(\sigma_i\xi(\mathbf{s}_i), \delta_i) = \tau(\sigma_i\mathbf{s}_i, \sigma_i) = (2\pi/\sigma_i)^{-1/2}\exp(\sigma_i\mathbf{s}_i^2/2)$. After observing their private signals all engage in repeated communications with their neighbors. Finally, we assume that each agents is trying to estimate the mean $m_{i,\theta} = \mathbb{E}_{i,\theta}\{\mathbf{s}_i\}$ of her private signal with as little variance as possible. Under the prescribed setting, Theorem 1 applies and the Bayesian heuristic update rules are affine with the coefficients as specified in the theorem with $n_i = 1$ and $\sigma_i = \delta_i$ for all $i$. In particular, if $\alpha_i = (0, \ldots, 0) \in \mathbb{R}^k$ and $\beta_i \to 0$ for all $i$, then $\epsilon_i = 0$ for all $i$ and the coefficients $T_{ij} = \sigma_j/\sum_{p \in \mathcal{N}_i}\sigma_p > 0$ specify a convex combination: $\sum_{j \in \mathcal{N}_i} T_{ij} = 1$ for all $i$.

---

[1] Some of the non-Bayesian update rules have the property that they resemble the replication of a first step of a Bayesian update from a common prior. For instance, DeMarzo, Vayanos and Zwiebel [21] interpret the weights in the DeGroot model as those assigned initially by rational agents to the noisy opinions of their neighbors based on their perceived precision. However, by repeatedly applying the same weights over and over again, the agents ignore the need to update these weights and to account for repetitions in their information sources (the so-called persuasion bias); as one of our main objectives, we formalize this setup as a Bayesian heuristic.

EXAMPLE 2 (POISSON SIGNALS WITH GAMMA BELIEFS). As the second example, suppose that each agent observe $n_i$ i.i.d. Poisson signals $\mathbf{s}_{i,p} : p \in [n_i]$ with mean $\delta_i\theta$, so that $\Theta = \mathcal{A}_i = (0, +\infty)$ for all $i \in [n]$. Moreover, we take each agent's prior to be a Gamma distribution with parameters $\alpha_i > 0$ and $\beta_i > 0$, denoted Gamma$(\alpha_i, \beta_i)$:

$$\nu_i(\theta) := \frac{d\mathcal{V}_i}{d\Lambda_1} = \frac{\beta_i^{\alpha_i}}{\Gamma(\alpha_i)}\theta^{\alpha_i - 1}e^{-\beta_i\theta},$$

for all $\theta \in (0, \infty)$ and each $i \in [n]$. Note that here $\eta(\theta) = \log\theta$, $\gamma(\eta(\theta)) = \exp(\eta(\theta)) = \theta$, $\kappa(\alpha_i, \beta_i) = \Gamma(\alpha_i)\beta_i^{-\alpha_i}$, $m_{i,\theta} = \delta_i\theta$, $\xi(\mathbf{s}_{i,p}) = \mathbf{s}_{i,p}$, $\sigma_i = 1$ and $\tau(\sigma_i\xi(\mathbf{s}_{i,p}), \delta_i) = \delta_i^{\mathbf{s}_{i,p}}/(\mathbf{s}_{i,p}!)$, for all $i, p$. This setting corresponds also to a case of Poisson observers with common rate $\theta$ and individual exposures $\delta_i, i \in [n]$, cf. [33, p. 54]. The posterior distribution over $\Theta$ after observing of the sum of $n_i$ Poisson mean $\delta_i\theta$ samples is again a Gamma distribution with updated (random) parameters $\sum_{p=1}^{n_i}\mathbf{s}_{i,p} + \alpha_i$ and $n_i\delta_i + \beta_i$, [33, pp. 52–53]. Using a quadratic utility $-(a - \delta_i\theta)^2$, the expected pay-off at time zero is maximized by the $\delta_i$-scaled mean of the posterior Gamma belief distribution [33, p. 587]: $\mathbf{a}_{i,0} = \delta_i(\sum_{p=1}^{n_i}\mathbf{s}_{i,p} + \alpha_i)/(n_i\delta_i + \beta_i)$. Given the neighboring information $\sum_{p=1}^{n_j}\mathbf{s}_{j,p} = (n_j + \beta_j\delta_j^{-1})\mathbf{a}_{j,0} - \alpha_j$, $\forall j \in \mathcal{N}_i$, agent $i$ can refine her belief into a Gamma distribution with parameters $\alpha_i + \sum_{j \in \mathcal{N}_i}[(n_j + \beta_j\delta_j^{-1})\mathbf{a}_{j,0} - \alpha_j]$ and $\beta_i + \sum_{j \in \mathcal{N}_i}n_j\delta_j$. The subsequent optimal action at time 1 and the resultant Bayesian heuristics are as claimed in Theorem 1 with $\sigma_i = 1$ for all $i \in [n]$. Here if we let $\alpha_i, \beta_i \to 0$ and $\delta_i = \delta > 0$ for all $i$, then $\epsilon_i = 0$ for all $i$ and the coefficients $T_{ij} = n_j/\sum_{p \in \mathcal{N}_i}n_p > 0$ again specify a convex combination: $\sum_{j \in \mathcal{N}_i}T_{ij} = 1$ for all $i$ as in the DeGroot model. In the following two subsections, we shall further explore this correspondence with the DeGroot updates and the implied asymptotic consensus amongst the agents.

**3.1. Linear updating and convergence.** In general, the constant terms $\epsilon_i$ in Theorem 1 depend on the neighboring prior parameters $\alpha_j$, $j \in \mathcal{N}_i \setminus \{i\}$ and can be non-zero. Accumulation of constant terms over time when $\rho(T) \geq 1$ prevents the action profiles from converging to any finite values or may cause them to oscillate indefinitely (depending upon the model parameters). However, if the prior parameters are vanishingly small, then the affine action updates in Theorem 1 reduce to linear update and $\epsilon_i = 0$. This requirement on the prior parameters is captured by our next assumption.

ASSUMPTION 2 (Non-informative priors). *For a member $\mathcal{V}(\cdot; \alpha, \beta)$ of the conjugate family $\mathcal{F}_{\gamma,\eta}$ we denote the limit $\lim_{\alpha_i, \beta_i \to 0} \mathcal{V}(\cdot; \alpha, \beta)$ by $\mathcal{V}_\varnothing(\cdot)$ and refer to it as the non-informative (and improper, if $\mathcal{V}_\varnothing(\cdot) \notin \mathcal{F}_{\gamma,\eta}$) prior.* [1] *All agents start from a common non-informative prior: $\mathcal{V}_i(\cdot) = \mathcal{V}_\varnothing(\cdot)$, $\forall i$.*

---

[1] Conjugate priors offer a technique for deriving the prior distributions based on the sample distribution (likelihood structures). However, in lack of any prior information it is impossible to justify their application on any subjective basis or to determine their associated parameters for any agent. Subsequently, the use of non-informative priors is suggested by Bayesian analysts and various techniques for selecting non-informative priors is explored in the literature [55]. Amongst the many proposed techniques for selecting non-informative priors, Jeffery's method sets its choice

As the name suggest non-informative priors do not inform the agent's action at time 0 and the optimal action is completely determined by the observed signal $\mathbf{s}_{i,p} : p \in [n_i]$ and its likelihood structure, parameterized by $\sigma_i$ and $\delta_i$. If we let $\alpha_i, \beta_i \to 0$ in the expressions of $T_{ij}$ and $\epsilon_i$ from Theorem 1, then the affine action updates reduce to linear combinations and the preceding corollary is immediate.

COROLLARY 1 (**Linear updating**). *Under the exponential family signal-utility structure (Assumption 1) with non-informative priors (Assumption 2); the Bayesian heuristics describe each updated action $\mathbf{a}_{i,t}$ as a linear combination of the neighboring actions $\mathbf{a}_{j,t-1}, j \in \mathcal{N}_i$: $\mathbf{a}_{i,t} = \sum_{j \in \mathcal{N}_i} T_{ij} \mathbf{a}_{j,t-1}$, where $T_{ij} = \delta_i \sigma_j n_j / (\sigma_i \sum_{p \in \mathcal{N}_i} n_p \delta_p)$.*

The action profiles under Corollary 1 evolve as a homogeneous positive linear discrete-time system: $\overline{\mathbf{a}}_{t+1} = (T \otimes I_k) \mathbf{a}_t$ and if the spectral radius of $T$ is strictly less than unity, then $\lim_{t \to \infty} \overline{\mathbf{a}}_t = \overline{0}$. For a strongly connected social network with $T_{ii} > 0$ for all $i$ the Perron-Frobenius theory [85, Theorems 1.5 and 1.7] implies that $T$ has a simple positive real eigenvalue equal to $\rho(T)$. Moreover, the left and right eigenspaces associated with $\rho(T)$ are both one-dimensional with the corresponding eigenvectors $\bar{l} = (l_1, \ldots, l_n)^T$ and $\bar{r} = (r_1, \ldots, r_n)^T$, uniquely satisfying $\|\bar{l}\|_2 = \|\bar{r}\|_2 = 1$, $l_i > 0$, $r_i > 0$, $\forall i$ and $\sum_{i=1}^n l_i r_i = 1$. The magnitude of any other eigenvalue of $T$ is strictly less than $\rho(T)$. If $\rho(T) = 1$, then $\lim_{t \to \infty} \overline{\mathbf{a}}_t = \lim_{t \to \infty} (T^t \otimes I_k) \overline{\mathbf{a}}_0 = (\bar{r}\bar{l}^T \otimes I_k) \overline{\mathbf{a}}_0$; in particular, the asymptotic action profile may not represent a consensus although every action converges to some point within the convex hull of the initial actions $\{\mathbf{a}_{i,0}, i \in [n]\}$. If $\rho(T) > 1$, then the linear discrete-time dynamics is unstable and the action profiles may increase or decrease without bound; pushing the decision outcome to extremes, we can associate $\rho(T) > 1$ to cases of polarizing group interactions.

**3.2. DeGroot updates, consensus and efficiency.** In order for the linear action updates in Corollary 1 to constitute a convex combination as in the DeGroot model,[1] we need to introduce some additional restrictions on the likelihood structure of the private signals.

ASSUMPTION 3 (**Locally balanced likelihoods**). *The likelihood structures given in (3) are called locally balanced if for all $i \in [n]$, $(\delta_i/\sigma_i) = (\sum_{j \in \mathcal{N}_i} \delta_j n_j) / \sum_{j \in \mathcal{N}_i} \sigma_j n_j$.*

Assumption 3 signifies a local balance property for the two exponential family parameters $\sigma_i$ and $\delta_i$ and across every neighborhood in the network. In particular, we need for the likelihood structures of every agent $i$ and her neighborhood to satisfy: $\delta_i \sum_{j \in \mathcal{N}_i} \sigma_j n_j = \sigma_i \sum_{j \in \mathcal{N}_i} \delta_j n_j$. Since parameters $\sigma_i$ and $\delta_i$ are both measures of accuracy or precision for private signals of agent $i$, the balance condition in Assumption 3 imply that the signal precisions are spread evenly over the agents; i.e. the quality of observations obey a rule of social balance such that no agent is in a

---

proportional to the square root of Fisher's information measure of the likelihood structure [81, Section 3.5.3], while Laplace's classical principle of insufficient reason favors equiprobability leading to priors which are uniform over the parameter space.

[1] The use of linear averaging rules for modeling opinion dynamics has a long history in mathematical sociology and social psychology [30]; their origins can be traced to French's seminal work on "A Formal Theory of Social Power" [29]. This was followed up by Harary's investigation of the mathematical properties of the averaging model, including the consensus criteria, and its relations to Markov chain theory [41]. This model was later generalized to belief exchange dynamics and popularized by DeGroot's seminal work [20] on linear opinion pools. In engineering literature, the possibility to achieve consensus in a distributed fashion (through local interactions and information exchanges between neighbors) is very desirable in a variety of applications such as load balancing [6], distributed detection and estimation [15, 95, 54], tracking [56], sensor networks and data fusion [98, 99], as well as distributed control and robotics networks [46, 64]. Early works on development of consensus algorithms originated in 1980s with the works of Tsitsiklis et.al [94] who propose a weighted average protocol based on a linear iterative approach for achieving consensus: each node repeatedly updates its value as a weighted linear combination of its own value and those received by its neighbors.

position of superiority to everyone else. Indeed, fixing $\delta_i = \delta$ for all $i$, the latter condition reduces to a harmonic property for the parameters $\sigma_i$, when viewed as a function of their respective nodes (cf. [61, Section 2.1] for the definition and properties of harmonic functions):

$$\sigma_i = \sum_{j \in \mathcal{N}_i} \frac{n_j \sigma_j}{\sum_{k \in \mathcal{N}_i} n_k}, \delta_i = \delta, \forall i. \tag{4}$$

However, in a strongly conceded social network (4) cannot hold true unless $\sigma_i$ is a constant: $\sigma_i = \sigma$ for all $i$. Similarly, when $\sigma_i = \sigma$ is a constant, then under Assumption 3 $\delta_i$ is spread as a harmonic function over the network nodes, and therefore can only take a constant value $\delta_i = \delta$ for all $i$, cf. [61, Section 2.1, Maximum Principle]. In particular, fixing either of the parameters $\sigma_i$ or $\delta_i$ for all agents and under the local balance condition in Assumption 3, it follows that the other parameter should be also fixed across the network; hence, the ratio $\sigma_i/\delta_i$ will be a constant for all $i$. Later when we consider the efficiency of consensus action we introduce a strengthening of Assumption 3, called globally balanced likelihood (cf. Assumption 4), where the ratio $\delta_i/\sigma_i$ should be a constant for all agents across the network. Examples 1 and 2 above provide two scenarios in which the preceding balancedness conditions may be satisfied: *(i)* having $\sigma_i = \delta_i$ for all $i$, as was the case with the Gaussian signals in Example 1, erasures that the likelihoods are globally balanced; *(ii)* all agents receiving i.i.d. signals from a common distribution in Examples 2 (Poisson signals with the common rate $\theta$ and common exposure $\delta$) makes a case for likelihoods being locally balanced.

THEOREM 2 (**DeGroot updating and consensus**). *Under the exponential family signal-utility structure (Assumption 1), with non-informative priors (Assumption 2) and locally balanced likelihoods (Assumption 3); the updated action* $\mathbf{a}_{i,t}$ *is a convex combination of the neighboring actions* $\mathbf{a}_{j,t-1}, j \in \mathcal{N}_i$: $\mathbf{a}_{i,t} = \sum_{j \in \mathcal{N}_i} T_{ij} \mathbf{a}_{j,t-1}, \sum_{j \in \mathcal{N}_i} T_{ij} = 1$ *for all* $i$. *Hence, in a strongly connected social network the action profiles converge to a consensus, and the consensus value is a convex combination of the initial actions* $\mathbf{a}_{i,0} : i \in [n]$.

In light of Theorem 2, it is of interest to know if the consensus action agrees with the minimum variance unbiased estimator of $m_{i,\theta}$ given all the observations of every agent across the network, i.e. whether the Bayesian heuristics *efficiently* aggregate all the information amongst the networked agents. Our next result addresses this question. For that to hold we need to introduce a strengthening of Assumption 3:

ASSUMPTION 4 (**Globally balanced likelihoods**). *The likelihood structures given in* (3) *are called globally balanced if for all* $i \in [n]$ *and some common constant* $C > 0$, $\delta_i/\sigma_i = C$.

In particular, under Assumption 4, $\sigma_i \delta_j = \sigma_j \delta_i$ for all $i, j$, and it follows that the local balance of likelihoods is automatically satisfied. We call the consensus action *efficient* if it coincides with the minimum variance unbiased estimator of $m_{i,\theta}$ for all $i$ and given all the observations of every agent across the network. Our next result indicates that global balance is a necessary condition for the agents to reach consensus on a globally optimal (efficient) action. To proceed, let the network graph structure be encoded by its adjacency matrix $A$ defined as $[A]_{ij} = 1 \iff (j,i) \in \mathcal{E}$, and $[A]_{ij} = 0$ otherwise. To express the conditions for efficiency of consensus, we need to consider the set of all agents who listen to the beliefs of a given agent $j$; we denote this set of agents by $\mathcal{N}_j^{out} := \{i \in [n] : [I + A]_{ij} = 1\}$ and refer to them as the out-neighborhood of agent $j$. This is in contrast to her neighborhood $\mathcal{N}_j$, which is the set of all agents whom she listens to. Both sets $\mathcal{N}_j$ and $\mathcal{N}_j^{out}$ include agent $j$ as a member.

THEOREM 3 (**(In-)Efficiency of consensus**). *Under the exponential family signal-utility structure (Assumption 1) and with non-informative priors (Assumption 2); in a strongly connected social network the agents achieve consensus at an efficient action if, and only if, the likelihoods are globally balanced and* $\sum_{p \in \mathcal{N}_j^{out}} n_p \delta_p = \sum_{p \in \mathcal{N}_i} n_p \delta_p$, *for all* $i$ *and* $j$. *The efficient consensus action is then given by* $\mathbf{a}^\star = \sum_{j=1}^{n} \left( \delta_j n_j \mathbf{a}_{j,0} / \sum_{p=1}^{n} n_p \delta_p \right)$.

Following our discussing of Assumption 3 and equation (4), we pointed out that if either of the two parameters $\sigma_i$ and $\delta_i$ that characterize the exponential family distribution of (3) are held fixed amongst the agents, then the harmonicity condition required for the local balancedness of the likelihoods implies that the other parameter is also fixed for all the agents. Therefore the local balancedness in many familiar cases (see Example 2) restricts the agents to observing i.i.d. signals: allowing heterogeneity only in the sample sizes, but not in the distribution of each sample. This special case is treated in our next corollary, where we also provide the simpler forms of the Bayesian heuristics and their linearity coefficients in the i.i.d. case:

COROLLARY 2 (**DeGroot learning with i.i.d. samples**). *Suppose that each agent $i \in [n]$ observes $n_i$ i.i.d. samples belonging to the same exponential family signal-utility structure (Assumption 1 with $\sigma_i = \sigma$ and $\delta_i = \delta$ for all i). If the agents have non-informative priors (Assumption 2) and the social network is strongly connected social network, then following the Bayesian heuristics agents update their action according to the linear combination: $\mathbf{a}_{i,t} = \sum_{j \in \mathcal{N}_i} T_{ij} \mathbf{a}_{j,t-1}$, where $T_{ij} = n_j / \sum_{p \in \mathcal{N}_i} n_p$, and reach a consensus. The consensus action is efficient if, and only if, $\sum_{p \in \mathcal{N}_j^{out}} n_p = \sum_{p \in \mathcal{N}_i} n_p$ for all i and j, and the efficient consensus action is given by $\mathbf{a}^\star = \sum_{j=1}^n \left( \mathbf{a}_{j,0} n_j / \sum_{p=1}^n n_p \right).$*

It is notable that the consensus value pinpointed by Theorem 2 does not necessarily agree with the MVUE of $m_{i,\theta}$ given all the private signals of all agents across the network; in other words, by following Bayesian heuristics agents may not aggregate all the initial data efficiently. As a simple example, consider the exponential family signal-utility structure with non-informative priors (Assumptions 1 and 2) and suppose that every agent observes an i.i.d. sample from a common distribution $\mathcal{L}(\cdot|\theta; 1, 1)$. In this case, the action updates proceed by simple iterative averaging: $\mathbf{a}_{i,t} = (1/|\mathcal{N}_i|) \sum_{j \in \mathcal{N}_i} \mathbf{a}_{j,t-1}$ for all $i \in [n]$ and any $t \in \mathbb{N}$. For an undirected graph $\mathcal{G}$ it is well-known that the asymptotic consensus action following simple iterative averaging is the degree-weighted average $\sum_{i=1}^n (deg(i)/|\mathcal{E}|) \mathbf{a}_{i,0}$, cf. [38, Section II.C]; and the consensus action is different form the global MVUE $\mathbf{a}^\star = (1/n) \sum_{i=1}^n \mathbf{a}_{i,0}$ unless the social network is a regular graph in which case, $deg(i) = d$ is fixed for all $i$, and $|\mathcal{E}| = n.d$.

REMARK 1 (**EFFICIENCY OF BALANCED REGULAR STRUCTURES**). In general, if we assume that all agents receive the same number of i.i.d. samples from the same distribution, then the condition for efficiency of consensus, $\sum_{p \in \mathcal{N}_j^{out}} n_p = \sum_{p \in \mathcal{N}_i} n_p$, is satisfied for balanced regular structures. In such highly symmetric structures, the number of outgoing and incoming links are the same for every node and equal to fixed number $d$.

Our results shed light on the deviations from the globally efficient actions, when consensus is being achieved through the Bayesian heuristics. This inefficiency of Bayesian heuristics in globally aggregating the observations can be attributed to the agents' naivety in inferring the sources of their information, and their inability to interpret the actions of their neighbors rationally, [38]; in particular, the more central agents tend to influence the asymptotic outcomes unfairly. This sensitivity to social structure is also due to the failure of agents to correct for the repetitions in the sources of the their information: agent $i$ may receive multiple copies that are all influenced by the same observations from a far way agent; however, she fails to correct for these repetition in the sources of her observations, leading to the co-called persuasion bias, [21].[1]

---

[1] Sobel [90] provides a theoretical framework to study the interplay between rationality and group decisions and points out the subsequent inefficiencies in information aggregation. The seminal work of Janis [50] provides various examples involving the American foreign policy in the mid-twentieth where the desire for harmony or conformity in the group have resulted in bad group decisions, a phenomenon that he coins *groupthink*. Various other works have looked at the choice shift toward more extreme options [23, 91] and group polarization [44, 83].

**4. Log-Linear belief updates.** When the state space $\Theta$ is finite, the action space is the probability simplex and the agents bear a quadratic utility that measures the distance between their action and the point mass on the true state, the communication structure between the agents is rich enough for them to reveal their beliefs at every time period. The Bayesian heuristics in this case lead to a log-linear updating of beliefs similar to what we analyze in [73, 75] under the Bayesian without recall model; the chief question of interest to the latter works is whether the agents, after being exposed to sequence of private observations and while communicating with each other, can learn the truth using the Bayesian without recall update rules. [1] The learning framework of [73, 74, 75, 78] in which agents have access to an stream of new observations is in contrast with the group decision model of this paper; the difference being in the fact that here the agents have a single initial observation and engage in group decision making to come up with the best decision that aggregates their individual private data with those of the other group members.

Consider an environment where the state space is a finite set of cardinality $m$ and agents take actions over the $(m-1)$-simplex of probability measures while trying to minimize their distance to a point mass on the true state. Specifically, let $\Theta = \{\theta_1, \ldots, \theta_m\}$, $\mathcal{A}_i = \Delta\Theta$ and for any probability measure $\mathcal{M}(\cdot) \in \mathcal{A}_i$ with probability mass function $\mu(\cdot) := d\mathcal{M}/d\mathcal{K}$, suppose that

$$u_i(\mathcal{M}, \theta_j) = -(1 - \mu(\theta_j))^2 - \sum_{\substack{k=1, \\ k \neq j}}^{m} \mu(\theta_k)^2.$$

Subsequently, $\mathcal{M}_{i,t} = \arg\max_{\mathcal{M} \in \Delta\Theta} \mathbb{E}_{i,t}\{u_i(\mathcal{M}, \theta)\}$, and the agents proceed by truthfully announcing their beliefs to each other at every time step. In particular, if we denote the belief probability mass functions $\nu_i(\cdot) := d\mathcal{V}_i/d\mathcal{K}$ and $\mu_{i,t}(\cdot) := d\mathcal{M}_{i,t}/d\mathcal{K}$ for all $t$, then we can follow the steps of [77] to derive the Bayesian heuristic $f_i$ in (2) by replicating the time-one Bayesian belief update for all future time-steps:

$$\boldsymbol{\mu}_{i,t}(\hat{\theta}) = \frac{\boldsymbol{\mu}_{i,t-1}(\hat{\theta}) \left( \prod_{j \in \mathcal{N}_i \setminus \{i\}} \frac{\boldsymbol{\mu}_{j,t-1}(\hat{\theta})}{\nu_j(\hat{\theta})} \right)}{\sum_{\tilde{\theta} \in \Theta} \boldsymbol{\mu}_{i,t-1}(\tilde{\theta}) \left( \prod_{j \in \mathcal{N}_i \setminus \{i\}} \frac{\boldsymbol{\mu}_{j,t-1}(\tilde{\theta})}{\nu_j(\tilde{\theta})} \right)}, \tag{5}$$

for all $\hat{\theta} \in \Theta$ and at any $t > 1$.[2,3]

---

[1] Naivety of agents in these cases impedes their ability to learn; except in simple social structures such as cycles or rooted trees (cf. [74]). Rahimian, Shahrampour and Jadbabaie [78] show that learning in social network with complex neighborhood structures can be achieved if agents choose a neighbor randomly at every round and restrict their belief update to the selected neighbor each time. In the literature on theory of learning in games [31], log-linear learning refers to a class of randomized strategies, where the probability of each action is proportional to an exponential of the difference between the utility of taking that action and the utility of the optimal choice. Such randomized strategies combine in a log-linear manner [14], and they have desirable convergence properties: under proper conditions, it can be shown that the limiting (stationary) distribution of action profiles is supported over the Nash equilibria of the game [62].

[2] In writing (5), every time agent $i$ regards each of her neighbors $j \in \mathcal{N}_i$ as having started from some prior belief $\nu_j(\cdot)$ and arrived at their currently reported belief $\boldsymbol{\mu}_{j,t-1}(\cdot)$ upon observing their private signals, hence rejecting any possibility of a past history, or learning and correlation between their neighbors. Such a rule is of course not the optimum Bayesian update of agent $i$ at any step $t > 1$, because the agent is not taking into account the complete observed history of beliefs and is instead, basing her inference entirely on the initial signals and the immediately observed beliefs.

[3] It is notable that the Bayesian heuristic in (5) has a log-linear structure. Geometric averaging and logarithmic opinion pools have a long history in Bayesian analysis and behavioral decision models [37, 84] and they can be also justified under specific behavioral assumptions [65]. The are also quite popular as a non-Bayesian update rule in engineering literature for addressing problems such as distributed detection and estimation [86, 79, 71, 59, 8]. In [8] the authors use

**4.1. An algebra of beliefs.**   Both the linear action updates studied in the previous chapter as well as the weighted majority update rules that arise in the binary case and are studied in [76, 75] have a familiar algebraic structure. It is instructive to develop similar structural properties for belief updates in (5) and over the space $\Delta\Theta$, i.e. the points of the standard $(m-1)$-simplex. Given two beliefs $\mu_1(\cdot)$ and $\mu_2(\cdot)$ over $\Theta$ we denote their "addition" as

$$\mu_1 \oplus \mu_2(\hat{\theta}) = \frac{\mu_1(\hat{\theta})\mu_2(\hat{\theta})}{\sum_{\tilde{\theta}\in\Theta}\mu_1(\tilde{\theta})\mu_2(\tilde{\theta})}.$$

Indeed, let $\Delta\Theta^o$ denote the $(m-1)$-simplex of probability measure over $\Theta$ after all the edges are excluded; $\Delta\Theta^o$ endowed with the $\oplus$ operation, constitutes a group (in the algebraic sense of the word). It is easy to verify that the uniform distribution $\bar{\mu}(\hat{\theta}) = 1/|\Theta|$ acts as the identity element for the group; in the sense that $\bar{\mu}\oplus\mu = \mu$ for all $\mu\in\Delta\Theta^o$, and given any such $\mu$ we can uniquely identify its inverse as follows:

$$\mu^{inv}(\hat{\theta}) = \frac{1/\mu(\hat{\theta})}{\sum_{\tilde{\theta}\in\Theta}1/\mu(\tilde{\theta})}.$$

Moreover, the group operation $\oplus$ is commutative and we can thus endow the abelian group $(\Delta\Theta^o, \oplus)$ with a subtraction operation:

$$\mu_1 \ominus \mu_2(\hat{\theta}) = \mu_1 \oplus \mu_2^{inv}(\hat{\theta}) = \frac{\mu_1(\hat{\theta})/\mu_2(\hat{\theta})}{\sum_{\tilde{\theta}\in\Theta}\mu_1(\tilde{\theta})/\mu_2(\tilde{\theta})}.$$

We are now in a position to rewrite the Bayesian heuristic for belief updates in terms of the group operations $\oplus$ and $\ominus$ over the simplex interior:

$$\boldsymbol{\mu}_{i,t} = \underset{j\in\mathcal{N}_i}{\oplus}\boldsymbol{\mu}_{j,t-1}\underset{j\in\mathcal{N}_i\setminus\{i\}}{\ominus}\nu_j.$$

The above belief update has a structure similar to the linear action updates studied in (1): the agents incorporate the beliefs of their neighbors while compensating for the neighboring priors to isolate the observational parts of the neighbors' reports. A key difference between the action and belief updates is in the fact that action updates studied in Section 3 are weighted in accordance with the observational ability of each neighbor, whereas the belief updates are not. Indeed, the quality of signals are already internalized in the reported beliefs of each neighbor; therefore there is no need to re-weight the reported beliefs when aggregating them.

a logarithmic opinion pool to combine the estimated posterior probability distributions in a Bayesian consensus filter; and show that as a result: the sum of KullbackLeibler divergences between the consensual probability distribution and the local posterior probability distributions is minimized. Minimizing the sum of KullbackLeibler divergences as a way to globally aggregate locally measured probability distributions is proposed in [11, 10] where the corresponding minimizer is dubbed the KullbackLeibler average. Similar interpretations of the log-linear update are offered in [70] as a gradient step for minimizing either the KullbackLeibler distance to the true distribution, or in [72] as a posterior incorporation of the most recent observations, such that the sum of KullbackLeibler distance to the local priors is minimized; indeed, the Bayes' rule itself has a product form and the Bayesian posterior can be characterized as the solution of an optimization problem involving the KullbackLeibler divergence to the prior distribution and subjected to the observed data [100]. In the past, we have investigated the implications of such log-linear behavior and properties of convergence and learning when agents are exposed to a stream of private observation [73, 74, 77, 78].

Given the abelian group structure we can further consider the "powers" of each element $\mu^2 = \mu \oplus \mu$ and so on; in general for each integer $n$ and any belief $\mu \in \Delta\Theta^o$, let the $n$-th power of $\mu$ be denoted by $n \odot \mu := \mu^n$, defined as follows:[1]

$$\mu^n(\hat{\theta}) = \frac{\mu^n(\hat{\theta})}{\sum_{\tilde{\theta} \in \Theta} \mu^n(\tilde{\theta})}.$$

Using the $\oplus$ and $\odot$ notations, as well as the adjacency matrix $A$ we get:

$$\boldsymbol{\mu}_{i,t+1} = \underset{j \in \mathcal{N}_i}{\oplus} \boldsymbol{\mu}_{j,t} \underset{j \in \mathcal{N}_i \setminus \{i\}}{\ominus} \nu_j = \underset{j \in [n]}{\oplus} \left([I+A]_{ij} \odot \boldsymbol{\mu}_{j,t}\right) \underset{j \in [n]}{\ominus} \left([A]_{ij} \odot \nu_j\right). \tag{9}$$

With some abuse of notation, we can concatenate the network beliefs at every time $t$ into a column vector $\boldsymbol{\mu}_t = (\boldsymbol{\mu}_{1,t}, \ldots, \boldsymbol{\mu}_{n,t})^T$ and similarly for the priors $\nu = (\nu_1, \ldots, \nu_n)^T$; thus (9) can be written in the vectorized format by using the matrix notation as follows:

$$\boldsymbol{\mu}_t = \{(I+A) \odot \boldsymbol{\mu}_{t-1}\} \ominus \{A \odot \nu\} \tag{10}$$

Iterating over $t$ and in the common matrix notation we obtain:

$$\boldsymbol{\mu}_t = \left\{(I+A)^t \odot \boldsymbol{\mu}_0\right\} \ominus \left\{\left(\sum_{\tau=0}^{t}(I+A)^{\tau}A\right) \odot \nu\right\}. \tag{11}$$

The above is key to understanding the evolution of beliefs under the Bayesian heuristics in (5), as we will explore next. In particular, when all agents have uniform priors $\nu_j = \bar{\mu}$ for all $j$, then (10) and (11) simplify as follows: $\boldsymbol{\mu}_t = (I+A) \odot \boldsymbol{\mu}_{t-1} = (I+A)^t \odot \boldsymbol{\mu}_0$. This assumption of a common uniform prior is the counterpart of Assumption 1 (non-informative priors) in Subsection 3.1, which paved the way for transition from affine action updates into linear ones. In the case of beliefs over a finite state space $\Theta$, the uniform prior $\bar{\mu}$ is non-informative. If all agents start form common uniform priors, the belief update in (5) simplifies as follows:

$$\boldsymbol{\mu}_{i,t}(\hat{\theta}) = \frac{\prod_{j \in \mathcal{N}_i} \boldsymbol{\mu}_{j,t-1}(\hat{\theta})}{\sum_{\tilde{\theta} \in \Theta} \prod_{j \in \mathcal{N}_i} \boldsymbol{\mu}_{j,t-1}(\tilde{\theta})}. \tag{12}$$

Our main focus in the next section is to understand how the individual beliefs evolve under (5), or (12) which is a spacial case of (5). The gist of our analysis is encapsulated in the group theoretic iterations: $\boldsymbol{\mu}_t = (I+A)^t \odot \boldsymbol{\mu}_0$, derived above for the common uniform priors case. In particular, our understanding of the increasing matrix powers $(I+A)^t$ plays a key role. When the network graph $\mathcal{G}$ is strongly connected, the matrix $I+A$ is primitive. The Perron-Frobenius theory [85, Theorems 1.5 and 1.7] implies that $I+A$ has a simple positive real eigenvalue equal to its spectral radius $\rho(I+A) = 1+\rho$, where we adopt the shorthand notation $\rho := \rho(A)$. Moreover, the left and right eigenspaces associated with this eigenvalue are both one-dimensional and the corresponding eigenvectors can be taken such that they both have strictly positive entries. The magnitude of any other eigenvalue of $I+A$ is strictly less than $1+\rho$. Hence, the eigenvalues of $I+A$ denoted by $\lambda_i(I+A)$, $i \in [n]$, can be ordered in their magnitudes as follows: $|\lambda_n(I+A)| \leq |\lambda_{n-1}(I+A)| \leq$

---

[1] This notation extends to all real numbers $n \in \mathbb{R}$, and it is easy to verify that the following distributive properties are satisfied:

$$n \odot (\mu_1 \oplus \mu_2) = (n \odot \mu_1) \oplus (n \odot \mu_2),$$
$$(m+n) \odot \mu_1 = (m \odot \mu_1) \oplus (n \odot \mu_1),$$
$$(m.n) \odot \mu_1 = m \odot (n \odot \mu_1),$$

for all $m, n \in \mathbb{R}$ and $\mu_1, \mu_2 \in \Delta\Theta^o$.

$\ldots < \lambda_1(I + A) = 1 + \rho$. Subsequently, we can employ the eigendecomposition of $(I + A)$ to analyze the behavior of $(I + A)^{t+1}$. Specifically, we can take a set of bi-orthonormal vectors $\bar{l}_i$, $\bar{r}_i$ as the left and right eigenvectors corresponding to the $i$th eigenvalue of $I + A$, satisfying: $\|\bar{l}_i\|_2 = \|\bar{r}_i\|_2 = 1$, $\bar{l}_i^T \bar{r}_i = 1$ for all $i$ and $\bar{l}_i^T \bar{r}_j = 0$, $i \neq j$; in particular, the left eigenspace associated with $\rho$ is one-dimensional with the corresponding eigenvector $\bar{l}_1 = \bar{\alpha} = (\alpha_1, \ldots, \alpha_n)^T$, uniquely satisfying $\sum_{i=1}^{n} \alpha_i = 1$, $\alpha_i > 0$, $\forall i \in [n]$, and $\bar{\alpha}^T A = (\rho + 1)\bar{\alpha}^T$. The entry $\alpha_i$ is called the centrality of agent $i$ and as the name suggests, it measures how central is the location of agent in the network. We can now use the spectral representation of $A$ to write [49, Section 6]:

$$(I + A)^t = (1 + \rho)^t \left( \bar{r}_1 \bar{\alpha}^T + \sum_{i=2}^{n} (\lambda_i(I + A)/(1 + \rho))^t \bar{r}_i \bar{l}_i^T \right). \tag{13}$$

Asymptotically, we get that all eigenvalues other than the Perron-Frobenius eigenvalue $1 + \rho$ are subdominant; hence, $(I + A)^t \to (1 + \rho)^t \bar{r}_1 \bar{\alpha}^T$ and $\boldsymbol{\mu}_t = (1 + \rho)^t \bar{r}_1 \bar{\alpha}^T \odot \boldsymbol{\mu}_0$ as $t \to \infty$; the latter holds true for the common uniform priors case and also in general, as we shall see next.

**4.2. Becoming certain about the group aggregate.** We begin our investigation of the evolution of beliefs under (12) by considering the optimal response (belief) of an agents who has been given access to the set of all private observations across the network; indeed, such a response can be achieved in practice if one follows Kahneman's advice and collect each individual's information privately before combining them or allowing the individuals to engage in public discussions [52, Chapter 23]. Starting from the uniform prior and after observing everybody's private data our aggregate belief about the truth state is given by the following implementation of the Bayes rule:

$$\boldsymbol{\mu}^{\star}(\hat{\theta}) = \frac{\prod_{j \in \mathcal{N}_i} \ell_j(\mathbf{s}_j | \hat{\theta})}{\sum_{\tilde{\theta} \in \Theta} \prod_{j \in \mathcal{N}_i} \ell_j(\mathbf{s}_j | \tilde{\theta})}. \tag{14}$$

Our next theorem describes the asymptotic outcome of the group decision process when the agents report their beliefs and follow the Bayesian heuristic (12) to aggregate them. The outcome indicated in Theorem 4 departs from the global optimum $\boldsymbol{\mu}^{\star}$ in two major respects. Firstly, the agents reach consensus on a belief that is supported over $\Theta^{\diamond} := \arg\max_{\tilde{\theta} \in \Theta} \sum_{i=1}^{n} \alpha_i \log(\ell_i(\mathbf{s}_i | \tilde{\theta}))$, as opposed to the global (network-wide) likelihood maximizer $\Theta^{\star} := \arg\max_{\tilde{\theta} \in \Theta} \boldsymbol{\mu}^{\star}(\hat{\theta}) = \arg\max_{\tilde{\theta} \in \Theta} \sum_{i=1}^{n} \log(\ell_i(\mathbf{s}_i | \tilde{\theta}))$; note that the signal log-likelihoods in the case of $\Theta^{\diamond}$ are weighted by the centralities, $\alpha_i$, of their respective nodes. Secondly, the consensus belief is concentrated uniformly over $\Theta^{\diamond}$, its support does not include the entire state space $\Theta$ and those states which score lower on the centrality-weighted likelihood scale are asymptotically rejected as a candidate for the truth state; in particular, if $\{\theta^{\diamond}\} = \Theta^{\diamond}$ is a singlton, then the agents effectively become certain about the truth state of $\theta^{\diamond}$, in spite of their essentially bounded aggregate information and in contrast with the rational (optimal) belief $\boldsymbol{\mu}^{\star}$ that is given by the Bayes rule in (14) and do not discredit or reject any of the less probable states. This unwarranted certainty in the face of limited aggregate data is a manifestation of the group polarization effect that derive the agent to more extreme beliefs, rejecting the possibility of any alternatives outside of $\Theta^{\diamond}$.

THEOREM 4 (**Certainty about the group aggregate**). *Following the Bayesian heuristic belief updates in* (5), $\lim_{t \to \infty} \boldsymbol{\mu}_{i,t}(\hat{\theta}) = 1/|\Theta^{\diamond}|$ *for all* $i \in [n]$ *and any* $\hat{\theta} \in \Theta^{\star}$, *where* $\Theta^{\diamond} := \arg\max_{\tilde{\theta} \in \Theta} \sum_{i=1}^{n} \alpha_i \log(\ell_i(\mathbf{s}_i | \tilde{\theta}))$. *In particular, if the sum of signal* log*-likelihoods weighted by node centralities is uniquely maximized by* $\theta^{\diamond}$, *i.e.* $\{\theta^{\diamond}\} = \Theta^{\diamond}$, *then* $\lim_{t \to \infty} \boldsymbol{\mu}_{i,t}(\theta^{\diamond}) = 1$ *almost surely for all* $i \in [n]$.

REMARK 2 (EFFICIENCY OF BALANCED REGULAR NETWORKS). The fact that log-likelihoods in $\Theta^\diamond$ are weighted by the node centralities is a source of inefficiency for the asymptotic outcome of the group decision process. This inefficiency is warded off in especially symmetric typologies, where in and out degrees of all nodes in the network are the same. In these so-called balanced regular digraphs, there is a fixed integer $d$ such that all agents receive reports from exactly $d$ agents, and also send their reports to some other $d$ agents; $d$-regular graphs are a special case, since all links are bidirectional and each agent sends her reports to and receive reports from the same $d$ agents. In such structures $\overline{\alpha} = (1/n)\mathbb{1}$ so that $\Theta^\star = \Theta^\diamond$ and the support of the consensus belief identifies the global maximum likelihood estimator (MLE); i.e. the maximum likelihood estimator of the unknown $\theta$, given the entire set of observations from all agents in the network.

**5. Conclusions.** We propose the Bayesian heuristics framework to address the problem of information aggregation and decision making in groups. Our model is consistent with the dual process theory of mind with one system developing the heuristics through deliberation and slow processing, and another system adopting the heuristics for fast and automatic decision making: once the time-one Bayesian update is developed, it is used as a heuristic for all future decision epochs. On the one hand, this model offers a behavioral foundation for non-Bayesian updating; in particular, linear action updates and log-linear belief updates. On the other hand, its deviation from the rational choice theory captures common fallacies of snap-judgments and history neglect that are observed in real life. Our behavioral method also complements the axiomatic approaches which investigate the structure of belief aggregation rules and require them to satisfy specific axioms such as label neutrality and imperfect recall, as well as independence or separability for log-linear and linear rules, respectively [65].

We showed that under a natural quadratic utility and for a wide class of distributions from the exponential family the Bayesian heuristics correspond to a minimum variance Bayes estimation with a known linear structure. If the agents have non-informative priors, and their signal structures satisfy certain homogeneity conditions, then these action updates constitute a convex combination as in the DeGroot model, where agents reach consensus on a point in the convex hull of their initial actions. In case of belief updates (when agents communicate their beliefs), we showed that the agents update their beliefs proportionally to the product of the self and neighboring beliefs. Subsequently, their beliefs converge to a consensus supported over a maximum likelihood set, where the signal likelihoods are weighted by the centralities of their respective agents.

Our results indicate certain deviations from the globally efficient outcomes, when consensus is being achieved through the Bayesian heuristics. This inefficiency of Bayesian heuristics in globally aggregating the observations is attributed to the agents' naivety in inferring the sources of their information, which makes them vulnerable to structural network influences: the share of centrally located agents in shaping the asymptotic outcome is more than what is warranted by the quality of their data. Another source of inefficiency is in the group polarization that arise as a result of repeated group interactions; in case of belief updates, this is manifested in the structure of the (asymptotic) consensus beliefs. The latter assigns zero probability to any alternative that scores lower than the maximum in the weighted likelihoods scale: the agents reject the possibility of less probable alternatives with certainty, in spite of their limited initial data. This overconfidence in the group aggregate and shift toward more extreme beliefs is a key indicator of group polarization and is demonstrated very well by the asymptotic outcome of the group decision process.

We pinpoint some key differences between the action and belief updates (linear and log-linear, respectively): the former are weighted updates, whereas the latter are unweighted symmetric updates. Accordingly, an agent weighs each neighbor's action differently and in accordance with the quality of their private signals (which are inferred from the actions). On the other hand, when communicating their beliefs the quality of each neighbor's signal is already internalized in their reported beliefs; hence, when incorporating her neighboring beliefs, an agent regards the reported beliefs of all her neighbors equally and symmetrically. Moreover, in the case of linear action updates the initial biases are amplified and accumulated in every iteration. Hence, the interactions of biased agents are very much dominated by their prior beliefs rather than their observations. This issue can push their choices to extremes, depending on the aggregate value of their initial biases. Therefore, if the Bayesian heuristics are to aggregate information from the observed actions satisfactorily, then it is necessary for the agents to be unbiased, i.e. they should hold non-informative priors about the state of the world and base their actions entirely on their observations. In contrast, when agents exchange beliefs with each other the multiplicative belief update can aggregate the observations, irrespective of the prior beliefs. The latter are asymptotically canceled; hence, multiplicative belief updates are robust to the influence of priors.

The Bayesian heuristics approach is strongly motivated by the behavioral processes that underlie human decision making. These processes often deviate from the predictions of the rational choice theory, and our investigation of the Bayesian heuristics highlights both the mechanisms for such deviations and their ramifications. In our ongoing research, we expand this behavioral approach by incorporating additional cognitive biases such as inattentiveness, and investigate how the decision processes are affected. On the one hand, the obtained insights highlight the value of educating the public about benefits of rational decision making and unbiased judgment, and how to avoid common cognitive errors when making decisions. On the other hand, by investigating the effect of cognitive biases, we can improve the practice of social and organizational policies, such that new designs can accommodate commonly observed biases, and work well in spite of them.

## Appendix A:   Proof of Theorem 1.

If agent $i$ starts from a prior belief $\mathcal{V}_i(\cdot) = \mathcal{V}(\cdot; \alpha_i, \beta_i) \in \mathcal{F}_{\gamma, \eta}$, then we can use the Bayes rule to verify that, cf. [81, Proposition 3.3.13], the Radon-Nikodym derivative of the Bayesian posterior of agent $i$ after observing $n_i$ samples $\mathbf{s}_{i,p} \in \mathcal{S}$, $p \in [n_i]$, with likelihood (3) is $\nu(\cdot; \alpha_i + \sigma_i \sum_{p=1}^{n_i} \xi(\mathbf{s}_{i,p}), \beta_i + n_i \delta_i)$, and in particular the Bayesian posterior at time zero belongs to the conjugate family $\mathcal{F}_{\gamma, \eta}$: $\mathbf{\mathcal{M}}_{i,0}(\cdot) = \mathcal{V}(\cdot; \alpha_i + \sigma_i \sum_{p=1}^{n_i} \xi(\mathbf{s}_{i,p}), \beta_i + n_i \delta_i)$.

Subject to the quadratic utility $u_i(a, \theta) = -(a - m_{i,\theta})^T(a - m_{i,\theta})$, the expected pay-off at any time time $t$ is maximized is by choosing [12, Lemma 1.4.1]:

$$\mathbf{a}_{i,t} = \mathbb{E}_{i,t}\{m_{i,\theta}\} := \int_{\theta \in \Theta} m_{i,\theta} \mathbf{\mathcal{M}}_{i,t}(d\theta),$$

which coincides with her minimum variance unbiased estimator (Bayes estimate) for $m_{i,\theta}$. The members of the conjugate family $\mathcal{F}_{\gamma, \eta}$ satisfy the following linearity property of the Bayes estimates that is key to our derivations.[1]

LEMMA 1 (**Proposition 3.3.14 of [81]**). *Let* $\zeta \in \mathbb{R}^k$ *be a parameter and suppose that the parameter space* $\Omega_\zeta$ *is an open set in* $\mathbb{R}^k$. *Suppose further that* $\zeta \in \Omega_\zeta$ *has the prior distribution* $\mathcal{W}(\cdot; \alpha, \beta)$ *with density* $\kappa'(\alpha, \beta)e^{\zeta^T \alpha - \beta \gamma(\zeta)}$ *w.r.t.* $\Lambda_k$ *where* $\kappa'(\alpha, \beta)$ *is the normalization constant. If* $\mathbf{s}' \in \mathcal{S}' \subset \mathbb{R}^k$ *is a random signal with distribution* $\mathcal{D}(\cdot; \zeta)$ *and density* $\tau'(\mathbf{s})e^{\zeta^T \mathbf{s} - \gamma'(\zeta)}$ *w.r.t.* $\Lambda_k$, *then*

$$\int_{\zeta \in \Omega_\zeta} \int_{s' \in \mathcal{S}'} s \mathcal{D}(ds; \zeta) \mathcal{W}(d\zeta; \alpha, \beta) = \frac{\alpha}{\beta}.$$

---

[1] In fact, such an affine mapping from the observations to the Bayes estimate characterizes the conjugate family $\mathcal{F}_{\gamma, \eta}$ and every member of this family can be uniquely identified from the constants of the affine transform [22].

Hence for any $\mathcal{V}(\cdot; \alpha, \beta) \in \mathcal{F}_{\gamma, \eta}$ we can write

$$
\begin{aligned}
\int_{\theta \in \Theta} m_{i,\theta} \mathcal{V}(d\theta; \alpha, \beta) &= \int_{\theta \in \Theta} \int_{s \in \mathcal{S}} \xi(s) \mathcal{L}(ds | \theta; \sigma, \delta) \mathcal{V}(d\theta; \alpha, \beta) \\
&= \int_{\theta \in \Theta} \left| \frac{\Lambda_k(\eta(d\theta))}{\mathcal{G}_\theta(d\theta)} \right| \frac{e^{\eta(\theta)^T \alpha - \beta \gamma(\eta(\theta))}}{\kappa(\alpha, \beta)} \mathcal{G}_\theta(d\theta) \times \\
&\quad \int_{s \in \mathcal{S}} \xi(s) \sigma \left| \frac{\Lambda_k(\xi(ds))}{\mathcal{G}_s(ds)} \right| \tau(\sigma\xi(s), \delta) e^{\sigma\eta(\theta)^T \xi(s) - \delta\gamma(\eta(\theta))} \mathcal{G}_s(ds) \\
&= \int_{\zeta \in \Omega_\theta} \left| \frac{\Lambda_k(\eta(d\theta))}{\mathcal{G}_\theta(d\theta)} \right| \frac{e^{\eta(\theta)^T \alpha - \beta\gamma(\eta(\theta))}}{\kappa(\alpha, \beta)} \mathcal{G}_\theta(d\theta) \times \\
&\quad \int_{s \in \mathcal{S}} \xi(s) \sigma \left| \frac{\Lambda_k(\xi(ds))}{\mathcal{G}_s(ds)} \right| \tau(\sigma\xi(s), \delta) e^{\sigma\eta(\theta)^T \xi(s) - \delta\gamma(\eta(\theta))} \mathcal{G}_s(ds) \\
&= \int_{\zeta \in \Omega_\eta} \frac{e^{\zeta^T \alpha - \frac{\beta}{\delta}\gamma'(\zeta)}}{\kappa(\alpha, \beta)} \Lambda_k(d\zeta) \int_{s' \in \mathcal{S}'} \frac{s'\tau'(s')}{\sigma} e^{\zeta^T s' - \gamma'(\zeta)} \Lambda_k(ds') \\
&= \frac{\alpha\delta}{\sigma\beta},
\end{aligned}
\tag{15}
$$

where in the penultimate equality we have employed the following change of variables: $\zeta = \eta(\theta)$, $s' = \sigma\xi(s)$, $\gamma'(\zeta) = \delta\gamma(\zeta)$, $\tau'(s') = \tau(s', \delta)$; and the last equality is a direct application of Lemma 1. In particular, given $\mathcal{M}_{i,0}(\cdot) = \mathcal{V}(\cdot; \alpha_i + \sigma_i \sum_{p=1}^{n_i} \xi(\mathbf{s}_{i,p}), \beta_i + n_i\delta_i)$, the expectation maximizing action at time zero coincides with:

$$
\mathbf{a}_{i,0} = \frac{\sum_{p=1}^{n_i} \xi(\mathbf{s}_{i,p}) + \sigma_i^{-1}\alpha_i}{n_i + \delta_i^{-1}\beta_i}.
\tag{16}
$$

Subsequently, following her observations of $\mathbf{a}_{j,0}$, $j \in \mathcal{N}_i$ and from her knowledge of her neighbor's priors and signal likelihood structure, agent $i$ infers the observed values of $\sum_{p=1}^{n_j} \xi(\mathbf{s}_{j,p})$ for all her neighbors. Hence, we get

$$
\sum_{p=1}^{n_j} \xi(\mathbf{s}_{j,p}) = (n_j + \delta_j^{-1}\beta_j)\mathbf{a}_{j,0} - \sigma_j^{-1}\alpha_j, \forall j \in \mathcal{N}_i.
\tag{17}
$$

The observations of agent $i$ are therefore augmented by the set of independent samples from her neighbors: $\{\sum_{p=1}^{n_j} \xi(\mathbf{s}_{j,p}) : j \in \mathcal{N}_i\}$, and her refined belief at time 1 is again a member of the conjugate family $\mathcal{F}_{\gamma, \eta}$ and is give by:

$$
\mathcal{M}_{i,1}(\cdot) = \mathcal{V}(\cdot; \alpha_i + \sum_{j \in \mathcal{N}_i} \sigma_j \sum_{p=1}^{n_j} \xi(\mathbf{s}_{j,p}), \beta_i + \sum_{j \in \mathcal{N}_i} n_j\delta_j).
$$

We can again invoke the linearity of the Bayes estimate for the conjugate family $\mathcal{F}_{\gamma, \eta}$ and the subsequent result in (15), to get that the expected pay-off maximizing action at time 1 is given by:

$$
\mathbf{a}_{i,1} = \frac{\delta_i \left( \alpha_i + \sum_{j \in \mathcal{N}_i} \sigma_j \sum_{p=1}^{n_j} \xi(\mathbf{s}_{j,p}) \right)}{\sigma_i \left( \beta_i + \sum_{j \in \mathcal{N}_i} n_j\delta_j \right)}.
\tag{18}
$$

Finally, we can use (17) to replace for the neighboring signals and derive the expression of the action update of agent $i$ at time 1 in terms of her own and the neighboring actions $\mathbf{a}_{j,0}$, $j \in \mathcal{N}_i$; leading to the expression of linear Bayesian heuristics as claimed in Theorem 1. $\qquad \square$

**Appendix B: Proof of Theorem 2.** The balancedness of likelihoods (Assumption 3) ensures that the coefficients of the linear conbination from Corollary 1 sum to one: $\sum_{j \in \mathcal{N}_i} T_{ij} = 1$, for all $i$; thus forming a convex combination as in the DeGroot model. Subsequently, the agents begin by setting $\mathbf{a}_{i,0} = \sum_{p=1}^{n_i} \xi(\mathbf{s}_{i,p})/n_i$ according to (16), and at every $t > 1$ they update their actions according to $\overline{\mathbf{a}}_t = T \overline{\mathbf{a}}_{t-1}$, where $\overline{\mathbf{a}}_t = (\mathbf{a}_{1,t}, \ldots, \mathbf{a}_{n,t})^T$ and $T$ is the $n \times n$ matrix whose $i,j$-th entry is $T_{ij}$. Next note from the analysis of convergence for DeGroot model, cf. [38, Propriotion 1], that for a strongly connected network $\mathcal{G}$ if it is aperiodic (meaning that one is the greatest common divisor of the lengths of all its circles; and it is the case for us, since the diagonal entries of $T$ are all non-zero), then $\lim_{\tau \to \infty} T^\tau = \mathbb{1}\overline{s}^T$, where $\overline{s} := (s_1, \ldots, s_n)^T$ is the unique left eigenvector associated with the unit eigenvalue of $T$ and satisfying $\sum_{i=1}^{n} s_i = 1$, $s_i > 0$, $\forall i$. Hence, starting from non-informative priors agents follow the DeGroot update and if $\mathcal{G}$ is also strongly connected, then they reach a consensus at $\overline{s}^T \overline{\mathbf{a}}_0 = \sum_{i=1}^{n} s_i (\sum_{p=1}^{n_i} \xi(\mathbf{s}_{i,p})/n_i)$. $\qquad \square$

**Appendix C: Proof of Theorem 3.** We begin by a lemma that determines the so-called global MVUE for each $i$, i.e. the MVUE of $m_{i,\theta}$ given all the observations of all agents across the network.

LEMMA 2 (**Global MVUE**). *Under the exponential family signal-utility structure (Assumption 1), the (global) MVUE of $m_{i,\theta}$ given the entire set of observations of all the agents across the network is given by:*

$$\mathbf{a}_i^\star = \frac{\delta_i \left( \alpha_i + \sum_{j=1}^{n} \sigma_j \sum_{p=1}^{n_j} \xi(\mathbf{s}_{j,p}) \right)}{\sigma_i \left( \beta_i + \sum_{j=1}^{n} n_j \delta_j \right)}. \tag{19}$$

*If we further impose non-informative priors (Assumption 2), then the global MVUE for each $i$ can be rewritten as*

$$\mathbf{a}_i^\star = \frac{\delta_i \left( \sum_{j=1}^{n} \sigma_j \sum_{p=1}^{n_j} \xi(\mathbf{s}_{j,p}) \right)}{\sigma_i \left( \sum_{j=1}^{n} n_j \delta_j \right)} = \frac{\delta_i}{\sigma_i} \sum_{j=1}^{n} \frac{\sigma_j n_j}{\sum_{p=1}^{n} n_p \delta_p} \mathbf{a}_{j,0}. \tag{20}$$

This lemma can be proved easily by following the same steps that lead to (18) to get (19); making the necessary substitutions under Assumption 2 yields (20). From (20), it is immediately clear that if some consensus action is to be the efficient estimator (global MVUE) for all agents $i \in [n]$, then we need $\delta_i \sigma_j = \sigma_i \delta_j$ for all $i, j$; hence, the global balance is indeed a necessary condition. Under this condition, the local balance of likelihoods (Assumption 3) is automatically satisfied and given non-informative priors Theorem 2 guarantees convergence to a consensus in strongly connected social network. Moreover, we can rewrite (20) as $\mathbf{a}_i^\star = \mathbf{a}^\star = (\sum_{j=1}^{n} \delta_j n_j \mathbf{a}_{j,0})/\sum_{p=1}^{n} n_p \delta_p$, for all $i$. Hence, if the consensus action ($\overline{s}^T \overline{\mathbf{a}}_0$ in the proof of Theorem 2, Appendix B) is to be efficient then we need $s_i = \delta_i n_i / \sum_{j=1}^{n} n_j \delta_j$ for all $i$; $\overline{s} = (s_1, \ldots, s_n)$ being the unique normalized left eigenvector associated with the unit eigenvalue of $T$: $\overline{s}^T T = \overline{s}^T$, as defined in Appendix B. Using $\delta_i \sigma_j = \sigma_i \delta_j$, we can also rewrite the coefficients $T_{ij}$ of the DeGroot update in Theorem 2 as $T_{ij} = \delta_j n_j / (\sum_{p \in \mathcal{N}_i} n_p \delta_p)$.

Therefore, by expanding the eigenvector condition $\overline{s}^T T = \overline{s}^T$ we obtain that in order for the consensus action $\overline{s}^T \overline{\mathbf{a}}_0$ to agree with the efficient consensus $\mathbf{a}^\star$, it is necessary and sufficient to have that for all $j$

$$\sum_{i=1}^{n} s_i T_{ij} = \sum_{i=1}^{n} \left( \frac{\delta_i n_i}{\sum_{j=1}^{n} \delta_j n_j} \right) \frac{\delta_j n_j [I + A]_{ij}}{\sum_{p \in \mathcal{N}_i} n_p \delta_p} = s_j = \frac{\delta_j n_j}{\sum_{j=1}^{n} \delta_j n_j}, \tag{21}$$

or equivalently,

$$\sum_{i:j \in \mathcal{N}_i} \frac{\delta_i n_i}{\sum_{p \in \mathcal{N}_i} n_p \delta_p} = \sum_{i \in \mathcal{N}_j^{out}} \frac{\delta_i n_i}{\sum_{p \in \mathcal{N}_i} n_p \delta_p} = 1, \tag{22}$$

for all $j$. Under the global balance condition (Assumption 4), $\delta_i \sigma_j = \delta_j \sigma_i$, the weights $T_{ij} = \delta_i n_j / (\sum_{p \in \mathcal{N}_i} n_p \delta_p)$ as given above, correspond to transition probabilities of a node-weighted random walk on the social network graph, cf. [16, Section 5]; where each node $i \in [n]$ is weighted by $w_i = n_i \delta_i$. Such a random walk is a special case of the more common type of random walks on weighted graphs where the edge weights determine the jump probabilities; indeed, if for any edge $(i, j) \in \mathcal{E}$ we set its weight equal to $w_{i,j} = w_i w_j$ then the random walk on the edge-weighted graph reduces to a random walk on the node-weighted graph with node weights $w_i, i \in [n]$. If the social network graph is undirected and connected (so that $w_{i,j} = w_{j,i}$ for all $i, j$), then the edge-weighted (whence also the node-weighted) random walks are time-reversible and their stationary distributions $(s_1, \ldots, s_n)^T$ can be calculated in closed form as follows [2, Section 3.2]:

$$s_i = \frac{\sum_{j \in \mathcal{N}_i} w_{i,j}}{\sum_{i=1}^{n} \sum_{j \in \mathcal{N}_i} w_{i,j}}. \tag{23}$$

In a node-weighted random walk we can replace $w_{i,j} = w_i w_j$ for all $j \in \mathcal{N}_i$ and (23) simplifies into

$$s_i = \frac{w_i \sum_{j \in \mathcal{N}_i} w_j}{\sum_{i=1}^{n} w_i \sum_{j \in \mathcal{N}_i} w_j}.$$

Similarly to (21), the consensus action will be efficient if and only if

$$s_i = \frac{w_i \sum_{j \in \mathcal{N}_i} w_j}{\sum_{i=1}^{n} w_i \sum_{j \in \mathcal{N}_i} w_j} = \frac{w_i}{\sum_{k=1}^{n} w_k}, \forall i,$$

or equivalently:

$$(\sum_{k=1}^{n} w_k) \sum_{j \in \mathcal{N}_i} w_j = \sum_{i=1}^{n} (w_i \sum_{j \in \mathcal{N}_i} w_j), \forall i,$$

which holds true only if $\sum_{j \in \mathcal{N}_i} w_j$ is a common constant that is the same for all agents, i.e. $\sum_{j \in \mathcal{N}_i} w_j = \sum_{j \in \mathcal{N}_i} \delta_j n_j = C' > 0$ for all $i \in [n]$. Next replacing in (22) yields that, in fact, $C' = \sum_{i \in \mathcal{N}_j^{out}} \delta_i n_i$ for all $j$, completing the proof for the conditions of efficiency. □

**Appendix D: Proof of Theorem 4.** We address the more general case of (5), which includes (12) as a special case (with common uniform priors). For any pair of states $\hat{\theta}$ and $\check{\theta}$, define the log ratio of beliefs, likelihoods, and priors as follows:

$$\begin{aligned}
\boldsymbol{\phi}_{i,t}(\hat{\theta}, \check{\theta}) &:= \log \left( \boldsymbol{\mu}_{i,t}(\hat{\theta}) / \boldsymbol{\mu}_{i,t}(\check{\theta}) \right), \\
\boldsymbol{\lambda}_i(\hat{\theta}, \check{\theta}) &:= \log \left( \ell_i(\mathbf{s}_i | \hat{\theta}) / \ell_i(\mathbf{s}_i | \check{\theta}) \right), \\
\gamma_i(\hat{\theta}, \check{\theta}) &:= \log \left( \nu_i(\hat{\theta}) / \nu_i(\check{\theta}) \right),
\end{aligned}$$

By concatenating the log-ratio statistics of the $n$ networked agents, we obtain the following three vectorizations for the log-ratio statistics:

$$\begin{aligned}
\overline{\boldsymbol{\phi}}_t(\hat{\theta}, \check{\theta}) &:= (\boldsymbol{\phi}_{1,t}(\hat{\theta}, \check{\theta}), \ldots, \boldsymbol{\phi}_{n,t}(\hat{\theta}, \check{\theta})), \\
\overline{\boldsymbol{\lambda}}(\hat{\theta}, \check{\theta}) &:= (\boldsymbol{\lambda}_1(\hat{\theta}, \check{\theta}), \ldots, \boldsymbol{\lambda}_n(\hat{\theta}, \check{\theta})), \\
\overline{\gamma}(\hat{\theta}, \check{\theta}) &:= (\gamma_1(\hat{\theta}, \check{\theta}), \ldots, \gamma_n(\hat{\theta}, \check{\theta})).
\end{aligned}$$

Using the above notation, we can rewrite the log-linear belief updates of (5) in a linearized vector format as shown below:

$$
\begin{aligned}
\overline{\boldsymbol{\phi}}_{t+1}(\hat{\theta},\check{\theta}) &= (I+A)\overline{\boldsymbol{\phi}}_t(\hat{\theta},\check{\theta}) - A\overline{\gamma}(\hat{\theta},\check{\theta}) \\
&= (I+A)^{t+1}\overline{\boldsymbol{\phi}}_0(\hat{\theta},\check{\theta}) - \sum_{\tau=0}^{t}(I+A)^{\tau}A\overline{\gamma}(\hat{\theta},\check{\theta}) \\
&= (I+A)^{t+1}\left(\overline{\boldsymbol{\lambda}}(\hat{\theta},\check{\theta}) + \overline{\gamma}(\hat{\theta},\check{\theta})\right) - \sum_{\tau=0}^{t}(I+A)^{\tau}A\overline{\gamma}(\hat{\theta},\check{\theta}) \\
&= (I+A)^{t+1}\overline{\boldsymbol{\lambda}}(\hat{\theta},\check{\theta}) + \left((I+A)^{t+1} - \sum_{\tau=0}^{t}(I+A)^{\tau}A\right)\overline{\gamma}(\hat{\theta},\check{\theta}).
\end{aligned}
$$

Next we use the spectral decomposition in (13) to obtain:[1]

$$
\begin{aligned}
\overline{\boldsymbol{\phi}}_{t+1}(\hat{\theta},\check{\theta}) &= (1+\rho)^{t+1}\overline{r}_1\boldsymbol{\Lambda}(\hat{\theta},\check{\theta}) + ((1+\rho)^{t+1} - \sum_{\tau=0}^{t}(1+\rho)^{\tau}\rho)\overline{r}_1\beta(\hat{\theta},\check{\theta}) + o((1+\rho)^{t+1}) \\
&= (1+\rho)^{t+1}\left(\overline{r}_1\boldsymbol{\Lambda}(\hat{\theta},\check{\theta}) + (1 - \sum_{\tau=0}^{t}(1+\rho)^{\tau-t-1}\rho)\overline{r}_1\beta(\hat{\theta},\check{\theta}) + o(1)\right) \\
&\to (1+\rho)^{t+1}\overline{r}_1\boldsymbol{\Lambda}(\hat{\theta},\check{\theta}),
\end{aligned}
\tag{24}
$$

where we adopt the following notations for the global log likelihood and prior ratio statistics: $\beta(\hat{\theta},\check{\theta}) := \overline{\alpha}^{T}\overline{\gamma}(\hat{\theta},\check{\theta})$ and $\boldsymbol{\Lambda}(\hat{\theta},\check{\theta}) := \overline{\alpha}^{T}\overline{\boldsymbol{\lambda}}(\hat{\theta},\check{\theta})$; furthermore, in calculation of the limit in the last step of (24) we use the geometric summation identity $\sum_{\tau=0}^{\infty}\rho(1+\rho)^{\tau-1} = 1$.

To proceed denote $\boldsymbol{\Lambda}(\hat{\theta}) := \sum_{i=1}^{n}\alpha_i\ell_i(\mathbf{s}_i|\hat{\theta})$ so that $\boldsymbol{\Lambda}(\hat{\theta},\check{\theta}) = \boldsymbol{\Lambda}(\hat{\theta}) - \boldsymbol{\Lambda}(\check{\theta})$. Since $\Theta^{\diamond}$ consists of the set of all maximizers of $\boldsymbol{\Lambda}(\hat{\theta})$, we have that $\boldsymbol{\Lambda}(\hat{\theta},\check{\theta}) < 0$ whenever $\check{\theta} \in \Theta^{\diamond}$ and $\hat{\theta} \notin \Theta^{\diamond}$. Next recall from (24) that $\overline{\boldsymbol{\phi}}_{t+1}(\hat{\theta},\check{\theta}) \to (1+\rho)^{t+1}\overline{r}_1\boldsymbol{\Lambda}(\hat{\theta},\check{\theta})$ and $\overline{r}_1$ is the right Perron-Frobenius eigenvector with all positive entries; hence, for all $\check{\theta} \in \Theta^{\diamond}$ and any $\hat{\theta}$, $\boldsymbol{\phi}_{i,t}(\hat{\theta},\check{\theta}) \to -\infty$ if $\hat{\theta} \notin \Theta^{\diamond}$ and $\boldsymbol{\phi}_{i,t}(\hat{\theta},\check{\theta}) = 0$ whenever $\hat{\theta} \in \Theta^{\diamond}$; or equivalently, $\boldsymbol{\mu}_{i,t}(\hat{\theta})/\boldsymbol{\mu}_{i,t}(\check{\theta}) \to 0$ for all $\hat{\theta} \notin \Theta^{\diamond}$, while $\boldsymbol{\mu}_{i,t}(\hat{\theta}) = \boldsymbol{\mu}_{i,t}(\check{\theta})$ for any $\hat{\theta} \in \Theta^{\diamond}$. The latter together with the fact that $\sum_{\theta \in \Theta}\boldsymbol{\mu}_{i,t}(\hat{\theta}) = 1$ for all $t$ implies that with probability one: $\lim_{t\to\infty}\boldsymbol{\mu}_{i,t}(\check{\theta}) = 1/|\Theta^{\diamond}|, \forall\check{\theta} \in \Theta^{\diamond}$ and $\lim_{t\to\infty}\boldsymbol{\mu}_{i,t}(\check{\theta}) = 0, \forall\check{\theta} \notin \Theta^{\diamond}$ as claimed in the Theorem. In the special case that $\Theta^{\diamond}$ is a singleton, $\{\theta^{\diamond}\} = \Theta^{\diamond}$, we get that $\lim_{t\to\infty}\boldsymbol{\mu}_{i,t}(\theta^{\diamond}) = 1$ almost surely for all $i \in [n]$.

**Acknowledgments.** This work was supported by ARO MURI W911NF-12-1-0509.

---

[1] Given two functions $f(\cdot)$ and $g(\cdot)$ we use the asymptotic notation $f(t) = o(g(t))$ to signify the relations $\lim_{t\to\infty}|f(t)/g(t)| = 0$.